\documentclass[twocolumn]{aastex62}

\usepackage{bm}
\usepackage{hyperref}
\usepackage{breakcites}
\usepackage{amsmath,amssymb,amsthm} 
\usepackage{url} 
\usepackage{float}
\usepackage{graphicx}
\usepackage{subfigure}
\usepackage{color}
\usepackage{lineno}

\usepackage{chngcntr}

\newcommand{\eqb}{\begin{eqnarray}}
\newcommand{\eqe}{\end{eqnarray}}
\newcommand{\bi}{\begin{itemize}}
\newcommand{\ei}{\end{itemize}}

\newcommand{\tab}[1]{Table~\ref{tab:#1}}

\newcommand{\angstrom}{\textup{\AA}}

\newcommand{\icnu}{IceCube-170922A}
\newcommand{\txs}{TXS~0506+056}
\newcommand{\swift}{\emph{Swift}}
\newcommand{\fermi}{\emph{Fermi}}
\newcommand{\maxi}{\emph{MAXI}}

\received{...}
\revised{...}
\accepted{...}
\submitjournal{ApJ}

\shorttitle{Longterm neutrino emission from \txs}
\shortauthors{Petropoulou et al.}

\begin{document}

\title{Multi-Epoch Modeling of \txs \, and Implications for Long-Term High-Energy Neutrino Emission}

\author{Maria~Petropoulou}
\affil{Department of Astrophysical Sciences, Princeton University,
Princeton, NJ 08544, USA}

\author{Kohta~Murase}
\affil{Department of Physics, Pennsylvania State University,
  University Park, PA 16802, USA}
\affil{Center for Particle \& Gravitational Astrophysics,
  Institute for Gravitation and the Cosmos,
  Pennsylvania State University,
  University~Park, PA 16802, USA}
\affil{Department of Astronomy \& Astrophysics,
  Pennsylvania State University,
  University Park, PA 16802, USA}
\affil{Center for Gravitational Physics, Yukawa Institute for
  Theoretical Physics, Kyoto, Kyoto 606-8502 Japan}

\author{Marcos~Santander}
\affil{Department of Physics and Astronomy, 
  University of Alabama, 
  Tuscaloosa, AL 35487, USA}

\author{Sara~Buson}
\affil{Institut f\"{u}r Th. Physik und Astrophysik, University of W\"{u}rzburg,  Emil-Fischer-Str. 31 D-97074, W\"{u}rzburg, Germany}
\affil{University of Maryland Baltimore County, 1000 Hilltop Circle, Baltimore, MD 21250, USA}

\author{Aaron~Tohuvavohu}
\affil{Department of Astronomy \& Astrophysics, University of Toronto, Toronto, Ontario, M5S 3H4, Canada}

\author{Taiki~Kawamuro} 
\affil{National Astronomical Observatory of Japan, Osawa, Mitaka, Tokyo 181-8588, Japan}

\author{Georgios~Vasilopoulos}
\affil{Department of Astronomy, Yale University, New Haven, CT 06520-8101, USA}

\author{Hiroshi~Negoro}
\affil{Department of Physics, Nihon University, 1-8 Kanda-Surugadai, Chiyoda-ku, Tokyo 101-8308, Japan}

\author{Yoshihiro~Ueda}
\affil{Department of Astronomy, Kyoto University, Kitashirakawa-Oiwake-cho, Sakyo-ku, Kyoto 606-8502, Japan} 

\author{Michael H. Siegel}
\affil{Department of Astronomy \& Astrophysics, Pennsylvania State University, University Park, PA 16802, USA}

\author{Azadeh~Keivani}
\affil{Department of Physics, Columbia University, New York, NY 10027}
\affil{Columbia Astrophysics Laboratory, Columbia University, New York, NY 10027}

\author{Nobuyuki~Kawai}
\affil{Department of Physics, Tokyo Institute of Technology, 2-12-1 Ookayama, Meguro-ku, Tokyo 152-8551, Japan}

\author{Apostolos~Mastichiadis}
\affil{Department of Physics, 
  National and Kapodistrian University of Athens, Panepistimiopolis, GR 15783 Zografos, Greece}
 
\author{Stavros~Dimitrakoudis}
\affil{Department of Physics, 
  University of Alberta, 
  Edmonton, Alberta T6G 2E1, Canada}


\begin{abstract}
The IceCube report of 
a $\sim 3.5\sigma$ excess of $13\pm5$ neutrino events in the direction of the blazar \txs\, in 2014--2015 and the 2017 detection of a high-energy neutrino, \icnu, during a  gamma-ray flare from the same blazar, have revived the interest in scenarios for neutrino production in blazars. We perform comprehensive analyses on the long-term electromagnetic emission of \txs \, using optical, X-ray, and gamma-ray data from the All-Sky Automated Survey for Supernovae (ASAS-SN), the \emph{Neil Gehrels Swift Observatory} (\emph{Swift}), the \emph{Monitor of  All-sky X-ray Image} (\maxi), and the  \emph{Fermi} Large Area Telescope (\emph{Fermi}-LAT). We also perform numerical modeling of the spectral energy distributions (SEDs) in four epochs prior to 2017 with contemporaneous gamma-ray and lower energy (optical and/or X-ray) data.
We find that the multi-epoch SEDs are consistent with a hybrid leptonic scenario, where the gamma-rays are produced in the blazar zone via external inverse Compton scattering of accelerated electrons, and high-energy neutrinos are produced via the photomeson production process of co-accelerated protons. The multi-epoch SEDs can be satisfactorily explained with the same jet parameters and variable external photon density and electron luminosity. Using the maximal neutrino flux derived for each epoch,
we put an upper limit of $\sim0.4-2$ on the muon neutrino number in ten years of IceCube observations. 
Our results are consistent with the \icnu \, detection, which can be explained as an upper fluctuation from the average neutrino rate expected from the source, but in strong tension with the 2014--2015 neutrino flare.
\end{abstract}


\section{Introduction}\label{sec:intro}
Active galactic nuclei (AGN), with relativistic jets powered by accretion onto their central supermassive black hole, are the most powerful persistent sources of electromagnetic radiation in the Universe, with bolometric luminosities of $\sim10^{43}-10^{48}$~erg s$^{-1}$ \citep[e.g.,][]{Ackermann2015}. An identifying property of blazars, a subclass of AGN with jets closely aligned to our line of sight \citep[][]{1993ARA&A..31..473A,Urry1995}, is their broadband (from radio wavelengths to GeV/TeV gamma-ray energies) variable emission, which can be significantly enhanced during  flares \citep[e.g.,][]{Aharonian2007, Fossati_2008, Ackermann2016, Ahnen2016}. 

On September 22 2017, the IceCube Neutrino Observatory detected a high-energy ($E_\nu\gtrsim290$ TeV) muon-track neutrino event (\icnu) in temporal and spatial coincidence with a multi-wavelength flare from a known blazar (\txs) at redshift $z=0.3365$ \citep{Ajello2014, Paiano:2018qeq}. This detection yielded the first ever $\sim3\sigma$ high-energy neutrino source association \citep{Aartsen2018blazar1}. 
A follow-up analysis of IceCube archival data revealed a past ``neutrino flare'' at a significance level of $\sim 3.5\sigma$ ($13\pm5$ signal events within $\sim$six months in 2014-2015) from the direction of \txs \, \citep{Aartsen2018blazar2}. The most probable energy for these neutrinos lies in the range $\sim 10 - 100\rm~TeV$, and the inferred isotropic-equivalent muon neutrino luminosity, if all signal events originated from \txs, is $\simeq 1.2 \times 10^{47}\rm~erg~s^{-1}$~\citep{Aartsen2018blazar2}. Notably, the neutrino flare was not accompanied by a gamma-ray flare or high flux in any other wavelength \citep[][but see \citealt{Padovani2018} for evidence of a 10 GeV gamma-ray flare]{Aartsen2018blazar2, Garrappa2019}.  

The reported association of \icnu \, with the 2017 multi-wavelength flare of \txs \, was studied in detail by several authors~\citep[e.g.,][]{Ansoldi2018, Keivani2018,Murase2018,Sahakyan2018, Liu2019,Gao2019,Cerruti2019}. Independently of  the details entering the theoretical calculations, most of the aforementioned studies concluded that $\lesssim 0.01-0.1$ muon neutrinos could have been emitted by \txs \, during its six month-long flare, if both neutrinos and the bulk of the blazar's electromagnetic radiation originated from the same region (henceforth, blazar zone). The predicted number of detected muon neutrinos, albeit low, is still consistent with the observation of \textit{one} neutrino from the 2017 flare of \txs~\citep{Aartsen2018blazar1,Strotjohann2019}. 

The inferred flux of the neutrino flare in 2014-2015 is well above the {\sl maximal} value set by cascade constraints in the context of single-zone scenarios of blazar emission \citep[for analytical estimates, see][]{Murase2018, Oikonomou2019}. The difficulty of explaining such a high neutrino flux from the same region that produces the non-thermal blazar emission mainly arises from the lack of enhanced electromagnetic activity, as revealed by available gamma-ray and optical data, during the period of the neutrino flare \citep[for detailed calculations, see][]{Reimer2019, Rodrigues2019}. 

If all neutrinos detected by IceCube from the direction of \txs \, are physically associated with this source, then they are suggestive of the following physical picture for the blazar: there should be at least two dissipation regions in the blazar jet, one responsible for the broadband emission with relatively low neutrino flux bound by the X-ray observations (blazar zone), and another one, more compact and likely transient, responsible for high neutrino fluxes and low GeV gamma-ray fluxes due to attenuation \citep[see discussions in][]{Reimer2019, Halzen2019}.

In this work, we estimate the long-term neutrino emission from the blazar zone of \txs. Although flares have been proposed as ideal periods for neutrino production \citep[e.g.,][]{Atoyan:2001ey, Halzen2005, Reimer2005, Dermer:2012rg, Dermer:2014vaa, Murase2018,Oikonomou:2019djc}, it is still likely that the neutrino flux from the blazar zone integrated over the lifetime of IceCube is high enough to explain the detection of neutrinos outside the six-month flaring period in 2017 \citep[see also][for Mrk 421]{Petropoulou:2016ujj}. 
Our  goal is to construct SEDs for different epochs prior to the 2017 multi-messenger flare, characterized by different flux levels in X-rays and gamma-rays, and determine the maximal neutrino flux from the blazar zone as a function of time. By converting these maximal neutrino fluxes into an expected number of neutrinos from \txs, we then provide a range for the number of muon tracks that can be seen in a future track sample in the IceCube data.
 
This paper is organized as follows. In Section \ref{sec:data} we present the data selection and analysis performed to construct multi-epoch SEDs of \txs. In Section \ref{sec:model} we briefly describe the scenario adopted for modeling the blazar SEDs. We present the results of the multi-epoch SED modeling in Section \ref{sec:SED-fits} and continue in Section \ref{sec:long} with our predictions for the long-term neutrino emission from \txs. We conclude in Sections \ref{sec:discussion} and \ref{sec:summary} with discussion of our results and a summary, focusing on uncertainties of the model, possible origins of the external radiation field required by our model as well as on the inferred jet power.

 Throughout the paper, we adopted a cosmology with $\Omega_M=0.3$, $\Omega_\Lambda=0.7$,  and $H_0=70$~km s$^{-1}$ Mpc$^{-1}$. The redshift of \txs \ \citep[$z=0.3365\pm0.0001$,][]{Paiano:2018qeq} corresponds then to a luminosity distance $d_L\simeq1774$~Mpc.

\section{Data selection and analysis}\label{sec:data}
We aim to characterize the broadband emission from the blazar zone of \txs~as a function of time. We focus our analysis on periods prior to the 2017 flare mainly for the following reasons: (i) we want to explore if the long-term neutrino emission prior to the 2017 flare can account for the detection of a high-energy neutrino like \icnu, and (ii) we want to include the period of the neutrino flare because of its special importance to theoretical models of neutrino production.  

For this purpose, we have selected archival multi-wavelength observations taken during three epochs outlined in Table~\ref{tab:Epochs}. The epochs have been selected based on the availability of contemporaneous gamma-ray and lower energy (optical and/or X-ray) data. We select a 159-day window for each epoch, which is consistent with the timescale reported by IceCube for the 2014-2015 neutrino flare~\citep{Aartsen2018blazar2}, as the latter offers the most secure estimate of a timescale for neutrino emission so far. 

\begin{figure*}
    \centering
    \includegraphics[width=\linewidth]{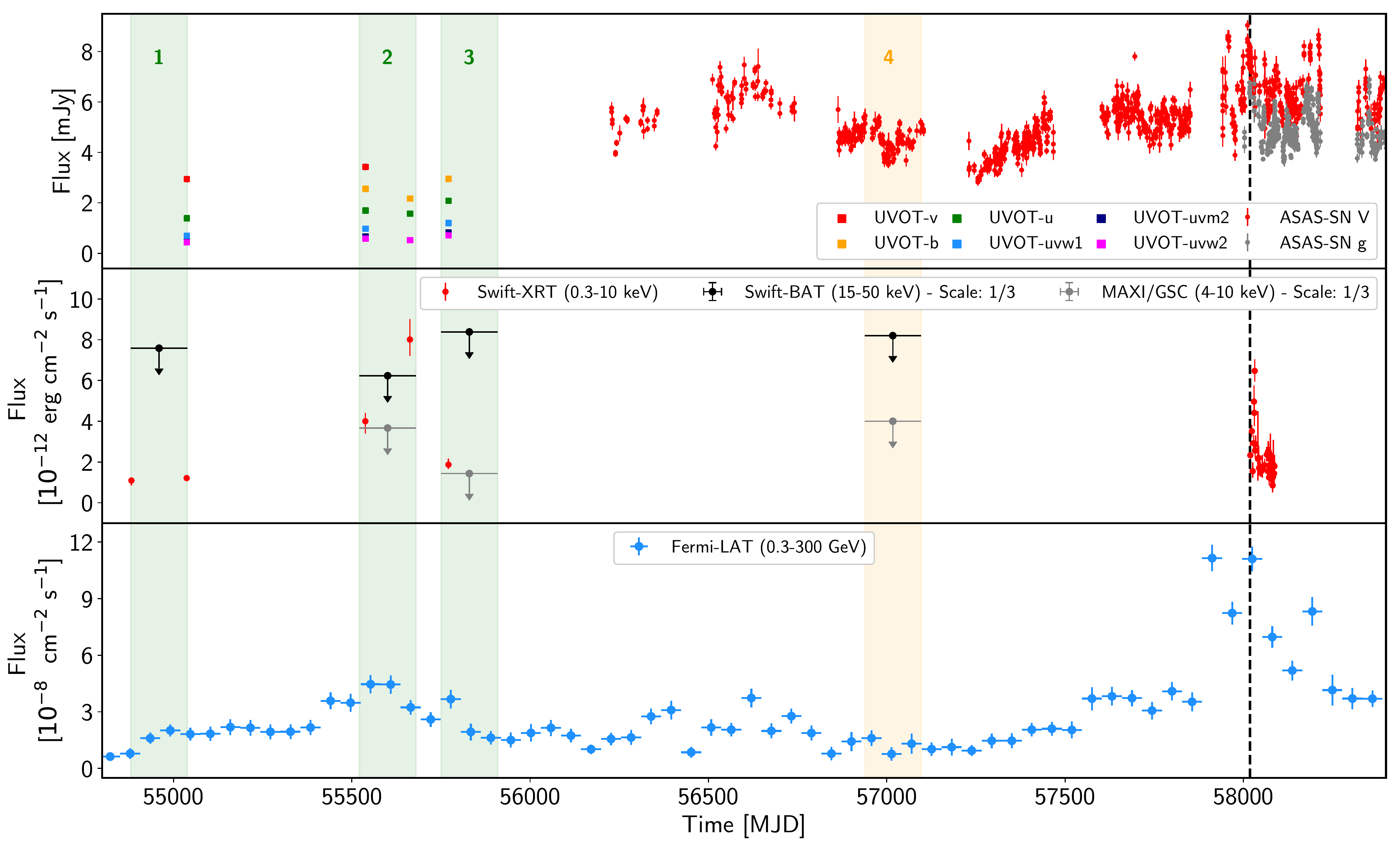}
    \caption{Multi-wavelength light curve of \txs \, composed of optical/UV data (not corrected for extinction) from ASAS-SN and \swift-UVOT  (top panel), X-ray data from \swift \, and \emph{MAXI}/GSC (middle panel), and gamma-ray data (in bins of 56.2 days) from \fermi-LAT (bottom panel). The shaded areas represent the epochs defined in Table~\ref{tab:Epochs} and used in our analysis. The black dashed line indicates the detection time of IceCube-170922A. \emph{Swift}-XRT observations after IceCube-170922A have been taken from~\cite{Keivani2018} and are shown for completeness. The \emph{MAXI}/GSC and \emph{Swift}-BAT upper limits have been scaled by a factor of 1/3 for better visibility. 
     }
\label{fig:LC}
\end{figure*}

The multi-wavelength light curve of the source, based on the analysis presented in the following subsections, is shown in Figure \ref{fig:LC} with the different epochs highlighted. Epochs 1 and 3 are characterized by a lower average gamma-ray and X-ray flux, while epoch 2 captures a period of enhanced emission in both X-ray and gamma-ray energy bands. 

\begin{deluxetable}{ccccc}
\centering
\tablecaption{Epoch definitions.
\label{tab:Epochs}}
\tablewidth{0pt}
\tablehead{
\colhead{Epoch} & \colhead{Start} & \colhead{End}  & \colhead{Start} & \colhead{End} 
\\ & [MJD] &[MJD] & [calendar] & [calendar]
}
\startdata
1 & 54880 & 55039 & 2009-02-18 & 2009-07-27  \\
2 & 55521 & 55680 & 2010-11-21 & 2011-04-29  \\ 
3 & 55750 & 55909 & 2011-07-08 & 2011-12-14 \\
4 & 56938 & 57096 & 2014-10-08 & 2015-03-15  \\
\enddata
\end{deluxetable}

\subsection{ASAS-SN}\label{sec:opt}
We use publicly available\footnote{\url{https://asas-sn.osu.edu/}} optical data from the All-Sky Automated Survey for Supernovae \citep[ASAS-SN;][]{Shappee2014, Kochanek2017}. The optical light curve from ASAS-SN is included in Figure \ref{fig:LC} (top panel). Data are available only for epoch 4. The time-averaged flux at the \textit{V}-band ($\nu\simeq5.4\times10^{14}$~Hz) is $4.4\pm0.4$~mJy or
$5.8\pm0.6$~mJy after correction for Galactic extinction using an $E(B-V)$ value of 0.108 from \cite{sf11} and following the extinction law of \cite{Fitzpatrick1999} with $R_{v} = 3.07$. 

\subsection{\swift-UVOT}\label{sec:uvot}
Observations taken with the Ultraviolet/Optical Telescope  (UVOT,~\citet{uvot2005}) onboard \swift \, were analyzed in \emph{image} mode (i.e., neglecting timing information) to characterize the optical/UV spectrum of the source. Five UVOT observations were identified in the epochs defined above, although the earliest one (ObsID 00038380001, taken on MJD 54882) was excluded as the frames show evidence of star trailing which could indicate some pointing instability. The UVOT periods, summarized in Table~\ref{tab:UVOT}, include exposures in several of the six broad-band UVOT filters (\textit{v}, \textit{b}, \textit{u}, \textit{uvw1}, \textit{uvm2}, \textit{uvw2}) described in \citet{poole2008,breeveld2011}. The values are also included in the light curve in Figure \ref{fig:LC} (top panel).

\begin{deluxetable*}{ccccc}
\centering
\tablecaption{\swift-UVOT photometry. The fluxes have been corrected for Galactic extinction. 
\label{tab:UVOT}}
\tablewidth{0pt}
\tablehead{\colhead{Epoch} & \colhead{ObsID} &\colhead{MJD}  & Filter  & \colhead{Flux [$10^{-11}$ erg cm$^{-2}$ s$^{-1}$]}   
  }
\startdata 
1 & 00038380002 & 55037 & \emph{uwm2} &  $1.61 \pm 0.07$ \\
1 & 00038380002 & 55037 & \emph{u} &  $1.91 \pm 0.07$ \\
1 & 00038380002 & 55037 & \emph{v} &  $2.22 \pm 0.1$ \\
1 & 00038380002 & 55037 & \emph{uvw1} & $1.54 \pm 0.05$ \\
1 & 00038380002 & 55037 & \emph{uvw2} &  $1.35 \pm 0.04$ \\
\hline
2 & 00040845001 & 55663 & \emph{b} &  $2.27 \pm 0.07$ \\
2 & 00040845001 & 55663 & \emph{u} &  $2.16 \pm 0.07$ \\
2 & 00040845001 & 55663 & \emph{uvw2} &  $1.61 \pm 0.05$ \\
\hline
2 & 00040845002 & 55771 & \emph{b} &  $3.09 \pm 0.12$ \\
2 & 00040845002 & 55771 & \emph{uvm2} &  $2.75 \pm 0.12$ \\
2 & 00040845002 & 55771 & \emph{u} &  $2.87 \pm 0.11$ \\
2 & 00040845002 & 55771 & \emph{uvw1} &  $2.66 \pm 0.12$ \\
2 & 00040845002 & 55771 & \emph{uvw2} &  $2.21 \pm 0.08$ \\
\hline
3 & 00041592001 & 55538 & \emph{b} & $2.67 \pm 0.08$ \\
3 & 00041592001 & 55538 & \emph{uvm2}  & $2.25 \pm 0.09$ \\
3 & 00041592001 & 55538 & \emph{u}  & $2.33 \pm 0.08$ \\
3 & 00041592001 & 55538 & \emph{v}  & $2.59 \pm 0.1$ \\
3 & 00041592001 & 55538 & \emph{uvw1}  & $2.17 \pm 0.07$ \\
3 & 00041592001 & 55538 & \emph{uvw2}  & $1.81 \pm 0.05$ \\
\enddata
\tablecomments{The central wavelengths (in \angstrom) of the \swift-UVOT filters are: 5468 (\emph{v}), 4392 (\emph{b}), 3465 (\emph{u}), 2600 (\emph{uvw1}), 2246 (\emph{uvm2}), and 1928 (\emph{uvw2}) \citep{poole2008,breeveld2011}.}
\end{deluxetable*}

The UVOT exposures were analyzed using the \texttt{uvotsource} tool included in {\sc heasoft} 6.25\footnote{\url{https://heasarc.gsfc.nasa.gov/docs/software/heasoft/}}. A 5-arcsec radius aperture was defined around the source, while a nearby 20-arcsec radius circular region with no evidence of faint sources was selected for background estimation. The data were calibrated using the latest UVOT \texttt{CALDB} files. 

The optical/UV fluxes were corrected for Galactic extinction using an $E(B-V)$ value of 0.108 from \cite{sf11}. Wavelength-dependent extinction coefficients were calculated at the central wavelength of each filter\footnote{\url{https://www.swift.ac.uk/analysis/uvot/filters.php}} following the extinction law of \cite{Fitzpatrick1999} with $R_{v} = 3.07$ using the York Extinction Solver~\citep{YES2004}. Extinction-corrected optical/UV flux values at the central wavelength for each filter are given in Table~\ref{tab:UVOT} and were used in the SED modeling shown in Figure \ref{fig:SEDs}.

\subsection{\swift-XRT}\label{sec:xrt}
We use X-ray data from the \emph{Neil Gehrels Swift Observatory} \citep{2004ApJ...611.1005G} X-ray telescope \citep[XRT, ][]{2005SSRv..120..165B}. \swift-XRT data products are available though the UK \swift \ Science Data Centre\footnote{\url{http://www.swift.ac.uk/user_objects/}}, and have been analyzed by using standard pipeline commands \citep{2007A&A...469..379E,2009MNRAS.397.1177E}. The pipeline produces light curves (i.e. count rate vs time) and spectral files in the $0.3-10$ keV energy band from all available observations. We identified five observations that fall within the periods of interest (see middle panel in Figure \ref{fig:LC}) and, for these, performed spectral fitting to constrain the spectral properties of \txs. Observations taken after the detection of IceCube-170922A are not included in this analysis, but are included in Figure \ref{fig:LC} for completeness.

The X-ray spectra were binned using at least one count per energy bin to allow the use of Cash statistics \citep{1979ApJ...228..939C}.
The spectral analysis of our data was performed with the {\sc xspec} fitting package V.~12.10.0 \citep{xspec}. 
All spectra were fitted with an absorbed power-law model, where the interstellar absorption was modeled using the {\tt tbnew} code \citep[{\tt tbabs} in the newest {\sc xspec} version]{wilms2000}, with Galactic abundances for elements heavier that He \citep{wilms2000} and appropriate atomic cross sections \citep{1996ApJ...465..487V}. 
First, we fitted individual observations with a model where all parameters were left free. Given the low statistics, the derived best-fit values were not significantly (i.e., beyond 3$\sigma$) different among individual observations.
We thus fitted all the individual data-sets simultaneously with the same model, using the same column density for all five observations and the same power-law slope for multiple observations within one epoch. The normalization of each of the five spectra was left as a free parameter, and the absorption was frozen at the expected value for the Galactic neutral atomic hydrogen column density $N_{\rm H}=1.16\times10^{21}$~cm$^{-2}$ \citep{HI4pi}.

The parameters of the best-fit power-law model (with their $1\sigma$ statistical errors) are presented in Table~\ref{tab:XRT} and the 0.3--10 keV fluxes are  displayed in Figure \ref{fig:LC} (middle panel).  The photon index $\Gamma$ derived by the fit is consistent with the values reported by \citet{Keivani2018}, who analyzed individual \swift-XRT observations after the \icnu~detection as part of the source monitoring program (see Figure 2 of the reference and Table~1 therein). By performing a simultaneous fit to \swift-XRT and \emph{NuSTAR} data, \citet{Keivani2018} found that the broadband X-ray spectrum of \txs\, is best fitted by a sum of two power-law components with best-fit photon indices of  $2.37\pm0.05$ (\swift-XRT band) and  $1.68\pm0.14$ for the hard part of the spectrum. 
The \swift-XRT data alone cannot put constraints on such model due to limited statistics and narrower energy range. We therefore did not test more sophisticated models than a simple power law.

\begin{deluxetable*}{ccccc}
\centering
\tablecaption{\swift-XRT  spectral fitting results in the $0.3-10$~keV energy at different epochs. 
\label{tab:XRT}}
\tablewidth{0pt}
\tablehead{\colhead{Epoch} & \colhead{ObsID} &\colhead{MJD}  & \colhead{Flux [$10^{-12}$ erg cm$^{-2}$ s$^{-1}$]} & \colhead{$\Gamma$}  
  }
\startdata 
1 & 00038380001 & 54882 &  1.09$^{+0.16}_{-0.24}$ & 2.30$^{+0.13}_{-0.14}$\\
1 & 00038380002 & 55037 &  1.21$^{+0.15}_{-0.11}$ & --\\
2\tablenotemark{\dag} & 00040845001 & 55663 &  8.01$^{+1.0}_{-0.8}$ & 2.40$^{+0.08}_{-0.17}$\\
2\tablenotemark{\ddag}  & 00041592001 & 55538  &4.0$^{+0.4}_{-0.6}$ &  --\\
3 & 00040845002 &  55771  & 1.87$^{+0.3}_{-0.21}$ & 2.4$^{+0.5}_{-0.3}$\\
\enddata
\tablecomments{Observations within the same epoch were fitted with the same power-law  slope. The C-statistic value of the combined fit is 270.65 for 293 degrees of freedom.  We report the $1\sigma$ statistical errors.
}
\tablenotetext{\dag}{High state.}
\tablenotetext{\ddag}{Low state.}
\end{deluxetable*}
\subsection{Swift-BAT}\label{sec:bat}
We created a mission-long (02/16/2005--08/25/2018) \swift-BAT light curve for TXS 0506+056 using the BAT Transient Monitor \citep{hans}, sensitive in the 15-50 keV bandpass, at 1, 2, 4, 8 and 16 day stacking intervals. The blazar was not detected on any time scale, allowing an average $3\sigma$ upper limit of 0.00053 cts~s$^{-1}$~cm$^{-2}$ to be set from the 16-day binned light curve. This corresponds to a flux upper limit of $3.03\times 10^{-11}$ erg~s$^{-1}$ cm$^{-2}$ throughout the entire period, assuming a power-law photon index of $\Gamma = 2.15$ \citep{tueller2010}. This upper limit is consistent with the non-detection of TXS 0506+056 to a depth of $7 \times 10^{-12}$ erg s$^{-1}$ cm$^{-2}$ in the BAT 105 month survey \citep{105month}, which includes all 8 spectral bands up to 195 keV and extends through August 2013. The $3 \sigma$ upper limits in the 15-50 keV band obtained for the periods of interest are summarized in Table~\ref{tab:BAT} and are included in Figure \ref{fig:LC} (middle panel).  These upper limits are constructed as the average 16-day bin limit during the relevant epochs, due to the issues with pattern noise.

\begin{deluxetable}{ccc}
\centering
\tablecaption{\swift-BAT $3\sigma$ upper limits in the $15-50$~keV energy at different epochs. 
\label{tab:BAT}}
\tablewidth{0pt}
\tablehead{
\colhead{Epoch} & \colhead{Count rate [cts cm$^{-2}$ s$^{-1}$]} & \colhead{Flux [erg~cm$^{-2}$ s$^{-1}$]}  
}
\startdata
1 & $3.98\times10^{-4}$ & $2.32\times10^{-11}$ \\
2 & $3.27\times10^{-4}$&  $1.94\times10^{-11}$ \\ 
3 & $4.4\times10^{-4}$&  $2.62\times10^{-11}$  \\
4 & $4.3\times10^{-4}$ &  $2.56\times10^{-11}$ \\
\enddata
\tablecomments{The count rate is converted to energy flux assuming a spectral index of $\Gamma = 1.7$, motivated by the hard spectrum of the 2017 flare in the \emph{NuSTAR} energy band  \citep{Aartsen2018blazar1,Keivani2018}.}
\end{deluxetable}

\subsection{MAXI}\label{sec:maxi}
We derived upper limits of soft X-ray (4--10 keV) fluxes by using data taken from the \maxi/Gas Slit Cameras \citep[GSCs; ][]{Matsuoka09,Mihara11}, which have been operated since 2009 August. Following previous analyses of \maxi/GSC data \citep[e.g., ][]{Kawamuro16a}, we performed 2D image fittings to observed images around TXS 0506+056 with a model composed of background and point spread function (PSF) models (see Figure 5 of \citealt{Hiroi13} for an example). We considered PSFs of all sources detected in the 7-year \maxi \, catalog \citep{Kawamuro18}, as well as TXS 0506+056 (which is not in the catalog). The PSFs were calculated using the \maxi \, simulator \citep{Eguchi09}  for $z = 0.3365$, by  assuming an absorbed power-law model with $N_{\rm H} = 1.16\times10^{21}$ cm$^{-2}$ and $\Gamma = 2.3$, motivated by the results of the \swift-XRT spectral analysis (see Section~\ref{sec:xrt}). The fluxes of all sources (in units of Crab) and the normalization of the background were left as free parameters. The positions of the 7-year \maxi \, catalog sources were fixed according to the results by \cite{Kawamuro18}, while that of TXS 0506+056 was set to its optical position. Note that only the \maxi/GSC data between 2009 August 13 and 2016 July 31 are currently examined and reprocessed so that we can perform the above fitting analyses. The source was not detected by \maxi/GSC in any of the considered epochs. No data are available for epoch 1 (MJD $54880-55039$) as it is prior to the operation of \maxi. The resulting $3\sigma$ upper limits are summarized in Table~\ref{tab:maxi} and are included in Figure \ref{fig:LC} (middle panel).

\begin{deluxetable}{ccc}
\centering
\tablecaption{\emph{MAXI}/GSC  $3\sigma$ upper limits in the $4-10$~keV  energy range  at different epochs.
\label{tab:maxi}}
\tablewidth{0pt}
\tablehead{
\colhead{Epoch} & \colhead{Flux [mCrab]} & \colhead{Flux [erg~cm$^{-2}$ s$^{-1}$]} 
}
\startdata
2 & 0.90 & $1.1\times10^{-11}$ \\ 
3 & 0.36 & $4.3\times10^{-12}$ \\
4 & 1.0 & $1.2\times10^{-11}$ \\
\enddata
\tablecomments{The flux conversion from Crab units to units of erg~cm$^{-2}$ s$^{-1}$ is made by assuming an absorbed power-law model with $N_{\rm H} = 1.16\times10^{21}$ cm$^{-2}$, $\Gamma = 2.3$, and $z = 0.3365$. }
\end{deluxetable}

\subsection{Fermi-LAT}\label{sec:fermi}
The \fermi\ Large Area Telescope (LAT) is a pair conversion telescope sensitive to gamma-rays in the 20 MeV to $>$300 GeV energy range~\citep{Atwood:2009ez}. In this work we analyze photon data collected by the LAT during four epochs outlined in Table~\ref{tab:Epochs}. Photons with energies between 100 MeV and 300 GeV that were detected within a $21^{\circ} \times 21^{\circ}$ square centered on the position of \txs~were included in the analysis, while photons detected with a zenith angle larger than 90$^{\circ}$ with respect to the spacecraft were discarded to reduce the Earth-limb gamma-ray contamination.

The contribution from isotropic and Galactic diffuse backgrounds was modeled using the parametrization provided in the \texttt{iso\_P8R2\_SOURCE\_V6\_v06.txt} and \texttt{gll\_iem\_v06.fits} files, respectively, leaving their normalization free to vary in the fit. Sources in the 3FGL catalog  within a radius of $20^{\circ}$ from the source position were included in the model with their spectral parameters fixed to their catalog values \citep{3FGL}, while the normalization and spectral indices for those within $3^{\circ}$ were allowed to vary freely during the spectral fit. The \txs~spectral fit was performed with a binned likelihood method using the \texttt{P8R2\_SOURCE\_V6} instrument response function. 

A power-law fit of the form $F(E) = F_0 (E/E_0)^{-\Gamma}$ was performed to characterize the spectral parameters of the source during each individual epoch over the entire 100 MeV to 300 GeV energy range. The best-fit parameters, flux normalization $F_0$ and photon index $\Gamma$, are given for each epoch in Table~\ref{tab:Fermi} at a normalization energy $E_0$ = 1.44 GeV. 
As an input to the SED modeling, the power-law fit was repeated in five independent energy bins with equal logarithmic spacing in the 100 MeV to 300 GeV for each epoch. In each bin, the flux normalization is left to vary freely while the photon index is kept at the best-fit value for the entire energy range. SED flux points with 68\% uncertainties are shown in Figure \ref{fig:SEDs} for those spectral bins with a test statistic\footnote{The test statistic (TS) is defined as the difference in the maximum likelihood of a model with and without the source. According to Wilks' theorem \citep{Wilks:1938} the TS distribution can be assumed to follow a $\chi^2$ distribution with one (two) degree of freedom for the power-law (log-parabola) spectral model \citep{Mattox:1996}.} (TS) larger than 4, which corresponds to a $2\sigma$ excess. Flux upper limits at the 95\% confidence level are shown otherwise.

The gamma-ray light curve of the source, shown in  Figure~\ref{fig:LC}, was built using the same response functions, quality cuts, window radius, and 3FGL parametrization as in the spectral analysis. The period between MJD 54682 and 58390, which corresponds to $\sim10$ years of \fermi's  operation, was divided into 66 equal-length bins, each about 56.2 days long. A power-law  spectral fit was performed in each time bin with the flux normalization and spectral index as free parameters. The best-fit parameters were used to calculate the photon flux in the 0.3--300 GeV energy range for each time bin of the light curve.

 We have also tested whether a log-parabolic model may provide a better description of the gamma-ray spectrum in each epoch. Our results indicate that the inclusion of the log-parabola fit parameters does not improve significantly ($<1\sigma$) the goodness of fit to the gamma-ray spectrum with respect to the power-law model for any epoch. While the log-parabola model is favored for the source in a long-term analysis \citep[as indicated in the 4FGL catalog,][]{4FGLarxiv} the limited exposure accumulated during the short time windows of each epoch limits the statistical power to discriminate between spectral models. For epoch 4 (i.e., the ``neutrino flare'' time interval) in particular, the LAT data do not provide convincing  evidence ($\lesssim2 \sigma$) of a spectral change. These results are consistent with those presented in other studies \citep{Padovani2018,Reimer2019,Garrappa2019}.

\begin{deluxetable*}{ccccc}
\centering
\tablecaption{Results from the analysis of \fermi-LAT data in the 0.1--300 GeV energy range. The table shows the gamma-ray flux, best-fit power-law spectral parameters, and test statistics (TS) values for different epochs. \label{tab:Fermi}}
\tablewidth{0pt}
\tablehead{
\colhead{Epoch} & \colhead{$F_{\mathrm{0.1-300 GeV}}$} & \colhead{$F_{0}$}  & \colhead{$\Gamma$} & \colhead{TS}  \\[-0.2cm]
 & \colhead{[erg cm$^{-2}$ s$^{-1}$]} & \colhead{[MeV$^{-1}$ cm$^{-2}$ s$^{-1}$]}  &  & }
\startdata
1 & $(5.0 \pm 0.6)\times 10^{-11}$ & $(2.1 \pm 0.2) \times 10^{-12}$ & $2.10 \pm 0.07$ & 310 \\ 
2 & $(1.5 \pm 0.1)\times 10^{-10}$ & $(5.8 \pm 0.3) \times 10^{-12}$ & $2.01 \pm 0.04$ & 1101 \\ 
3 & $(8.7 \pm 0.9)\times 10^{-11}$ & $(3.6 \pm 0.3) \times 10^{-12}$ & $2.09 \pm 0.07$  & 440 \\
4 & $(4.8 \pm 1.5)\times 10^{-11}$ &  $(1.5 \pm 0.3) \times 10^{-12}$ & $1.89 \pm 0.13$ & 117 \\ 
\enddata
\end{deluxetable*}

\section{Model description}\label{sec:model}
We adopt the standard one-zone model for blazar emission, according to which, the blazar SED (from infrared wavelengths to gamma-ray energies) is explained by synchrotron and inverse Compton processes of accelerated (henceforth, primary) electrons that are injected in a localized region, acting as the radiation zone of the blazar jet~\citep{maraschietal92,Dermer1992, Dermer1993,sikora94}. A population of relativistic protons, which is necessary for the production of high-energy neutrinos via the photomeson production process \citep[e.g.,][]{Sikora1987}, is also assumed to be injected in the same region. Neutrons, which are also a by-product of the photomeson production process~\citep[e.g.,][]{Kirk1989, Atoyan:2002gu, Dermer:2012rg}, can escape almost unimpeded from the blazar zone for typical parameters,  as those used in this work (see Section~\ref{sec:SED-fits}).

In addition to photomeson interactions, protons lose energy by synchrotron radiation and Bethe-Heitler pair production. These processes, together with photon-photon  pair production, are an important source of secondary electron-positron pairs. The latter lose energy by synchrotron radiation and inverse Compton scattering, and can contribute to the broadband photon emission. The model considers both synchrotron photons and external (to the jet) photons as seeds/targets for inverse Compton scattering, photohadronic ($p\gamma$) interactions, and photon-photon pair production. We do not specify the origin of the external radiation field in an attempt to be as model-independent as possible. We discuss possible origins for the external radiation field in Section~\ref{sec:discussion}.
Our working hypothesis is that the electromagnetic radiation produced by the secondaries from $p\gamma$ interactions is hidden by the primary leptonic emission and is not directly observable (for more details, see Section~3.2 in \citet{Keivani2018} and Section~3.1 in \citet{Murase2018}). 

The interplay of the aforementioned physical processes, which governs the evolution of the particle energy distributions, can be described by a set of time-dependent coupled integrodifferential equations. With this numerical scheme, energy is conserved in a self-consistent way, since all the energy gained by one particle species has to come from an equal amount of energy lost by another particle species. 
To simultaneously solve the coupled kinetic equations for all particle types we use the time-dependent code\footnote{The code uses the routine {\tt d02ejf} from the NAG Fortran library, which integrates a stiff system of first-order ordinary differential equations over an interval with suitable initial conditions.} described in \citet{DMPR12}. Photomeson production processes are modeled using the results of the Monte Carlo event generator {\sc sophia}~\citep{SOPHIA2000}, while the Bethe-Heitler pair production is similarly modeled with the Monte Carlo results of \citet{Protheroe1996} and \citet{mastetal05}. 
The only particles that are not modeled with kinetic equations are muons, pions, and kaons~\citep{DPM14,petroetal14}; their energy losses can be safely ignored for the parameter values relevant to this study (see also \citealt{Murase:2007yt} for numerical calculations where the kinetic equations for these particles are explicitly solved). The adopted numerical scheme is ideal for studying the development of electromagnetic cascades in the source in both linear and non-linear regimes\footnote{As long as the energy density of the secondary photons is lower than that of the synchrotron photons from primary electrons (and/or external radiation fields), the cascade is considered to be linear, i.e., the interactions between secondary pairs and photons are negligible. If this is not true, the cascade is characterized as non-linear.} \citep[see, e.g.,][]{petro2014, PM18, petro_yuan19}.
\begin{table*}[t]
\caption{Physical parameters (description, symbol, and units) used in the single-zone hybrid leptonic model of blazar emission.}
\begin{center}
\begin{tabular}{ccc}
\hline 
Parameter & Symbol & Unit [in cgs]\\
\hline 
Doppler factor & $\delta$ & n/a \\
Bulk Lorentz factor & $\Gamma$ & n/a \\
Magnetic field strength & $B^\prime$ & G \\
Blob radius & $R^\prime$ & cm \\
Electron injection luminosity & $L_e^\prime$ & erg s$^{-1}$ \\
Minimum electron Lorentz factor & $\gamma^\prime_{e, \min}$ & n/a \\
Maximum electron Lorentz factor & $\gamma^\prime_{e, \max}$ & n/a \\
Break electron Lorentz factor & $\gamma^\prime_{e, br}$ & n/a \\
Power-law electron index below the break &  $s_{e,1}$ & n/a \\
Power-law electron index above the break &  $s_{e,2}$ & n/a \\
Proton injection luminosity & $L_p^\prime$ & erg s$^{-1}$ \\
Minimum proton Lorentz factor & $\gamma^\prime_{p, \min}$ & n/a \\
Maximum  proton Lorentz factor & $\gamma^\prime_{p, \max}$ & n/a \\
Power-law proton index &  $s_{p}$ & n/a \\ 
Energy density of external radiation & $u^\prime_{\rm ext}$ & erg cm$^{-3}$ \\
Effective temperature of gray-body external radiation & $T_{\rm ext}^\prime$ & K \\
\hline
\end{tabular}
\end{center}
\tablecomments{Primed quantities are measured in the jet (blob) comoving frame. Unprimed quantities are measured in the observer's rest frame, unless stated otherwise. Parameters describing the relativistic particle distributions refer to their properties at injection, i.e., before modification due to cooling.}
\label{tab:definitions}
\end{table*} 

\begin{deluxetable*}{l cccccc} 
\centering
\tablecaption{Parameter values of the hybrid leptonic model for the multi-epoch observations of \txs. \label{tab:fit}}
\tablewidth{0pt}
\tablehead{
\colhead{Parameter} & \multicolumn{6}{c}{Value} \\
          & \colhead{Epoch~1} & \multicolumn{2}{c}{Epoch~2} & \colhead{Epoch~3} & \colhead{Epoch~4} & \colhead{2017 flare\tablenotemark{\dag}}\\
          & \colhead{(MJD~$54880-55039$)} & \multicolumn{2}{c}{(MJD~$55521-55680$)} & \colhead{(MJD~$55750-55909$)} & 
          \colhead{(MJD~$56938-57096$)} &
          \colhead{(MJD~$58003-58033$)}
          }
\startdata
& & XRT low & XRT high & \multicolumn{3}{c}{}\\
\hline
$u_{\rm ext}^{\prime}$ [erg cm$^{-3}$] &  $1.1\times10^{-2}$  &  \multicolumn{2}{c}{ $2.9\times10^{-2}$} &  $1.5\times10^{-2}$ & $6.6\times10^{-3}$ & $3.3\times10^{-2}$\\
$L^\prime_{e}$ [erg s$^{-1}$] & $8.6\times10^{41}$ &  \multicolumn{2}{c}{$1.1\times 10^{42}$} & $10^{42}$ & $8.8\times 10^{41}$ & $2.2\times 10^{42}$\\
$s_{e,2}$ & 3.6 & 3.2 & 2.9 & 3.7 & 3.6 & 3.6\\
$\gamma^\prime_{e, b}$ & $5\times10^3$ & \multicolumn{2}{c}{$5\times10^3$} & $8\times 10^3$ & $8\times 10^3$  & $5\times 10^3$ \\
$\gamma^\prime_{e, \rm max}$ & $3.2\times10^5$ &  \multicolumn{2}{c}{$3.2\times10^5$} & $10^5$ & $10^5$ & $10^5$  \\ 
$L_{p}^{\prime (\rm max)}$ [erg~s$^{-1}$] & $4\times10^{44}$& $5.4\times10^{44}$ & $1.3\times10^{45}$ &  $1.7\times10^{45}$ & $2.7\times10^{45}$ & $5.4\times10^{44}$\\ 
$L_{p,j}^{(\max)}$\tablenotemark{\ddag} [erg~s$^{-1}$] & $8.8\times10^{46}$& $4.8\times10^{47}$ & $2.9\times10^{47}$ &  $3.8\times10^{47}$ & $6\times10^{47}$ & $1.2\times10^{47}$\\
\enddata
\tablecomments{Other parameters used in the model, but kept fixed at all epochs, are: $\delta=24.2$, $B^\prime=0.4$~G, $R^\prime=10^{17}$~cm, $T_{\rm ext}^\prime=3\times10^5$~K ($\epsilon^\prime_{\rm ext}\approx3k_BT_{\rm ext}^\prime\simeq 78$~eV), $\gamma^\prime_{e, \min}=1$, $s_{e,1}=1.9$, $s_p=2$, $\gamma^\prime_{p,\min}=1$, and $\gamma^\prime_{p, \rm max}=1.6\times10^7$. For an explanation of the model parameters, see Table~\ref{tab:definitions}.}
\tablenotetext{\dag}{For a direct comparison to the modeling results of the 2017 flare of \txs \,  \citep{Keivani2018}, we list the parameters of the hybrid leptonic model LMBB2b (see also Table~7 of the reference).}
\tablenotetext{\ddag}{Absolute power  of a two-sided jet in protons defined as: $L_{p,j}=2\pi R^{\prime 2} c \Gamma^2 u^{\prime}_p \approx (3/2) \Gamma^2 L_p^{\prime}$, where $\Gamma \approx \delta/2$.}
\end{deluxetable*}

\section{Multi-epoch SED modeling}\label{sec:SED-fits}
Table~\ref{tab:definitions} summarizes the parameters used in the single-zone hybrid leptonic model of blazar emission, as outlined in the previous section. The parameters that describe the source (i.e., Doppler factor $\delta$, comoving magnetic field strength $B^\prime$ and blob radius $R^\prime$) as well as these of accelerated particle distributions (e.g., comoving injection electron luminosity $L_{e}^\prime$) can often be constrained by multi-wavelength data \citep{1996ApJ...470L..89T,mastkirk97,1998A&A...333..452K,2000ApJ...536..729L}. Instead of performing a blind parameter space search, we use as our starting point the parameter set of one of the hybrid leptonic models for the 2017 flare of \txs \ presented in \citet{Keivani2018}, which is the most optimistic in terms of neutrino production from the blazar zone \citep[see model LMBB2b in Table~7 and Figure 5 in][]{Keivani2018}. This parameter set will serve as our baseline model in the search for parameter values that describe the multi-epoch SEDs.

Differences in the observed multi-epoch SEDs can be explained by varying the parameter values of the baseline model for the 2017 flare. In principle, all model parameters are allowed to vary, thus leading to different physical descriptions of the same data-set. In what follows, we search for the simplest physical representation of the data, namely the one that can be achieved (within the adopted framework) by varying as few parameters as possible with respect to their baseline values.

 Assuming that the blazar zone is located at a fixed distance from the black hole, it is reasonable to consider that the properties of the emitting region (i.e., $R^\prime$, $B^\prime$, and $\delta$) are approximately constant in time. In this case, the only model parameters that are allowed to vary in order to explain the differences in the observed multi-epoch SEDs are the properties of the accelerated electrons and of the external photon field.
To minimize even further the number of free parameters, the photon energy spectrum of the external photon field is modeled as a gray body, which can be fully described with only two free parameters: its characteristic temperature\footnote{The corresponding average photon energy is $\epsilon_{\rm ext}^\prime\approx 3 k_B T^\prime_{\rm ext}$.} $T_{\rm ext}^\prime$ and energy density $u^\prime_{\rm ext}$ as measured in the rest frame of the emitting region. We also neglect any angular dependencies of the external radiation field, which for simplicity is assumed to be isotropic in the rest frame of the jet \citep[for anisotropic radiation fields, see, e.g.,][]{Protheroe1992, Dermer1993, Finke2016}.

 To determine the parameter values needed to explain the multi-epoch SEDs, we compute for each parameter set a steady-state\footnote{For each parameter set, we evolve the system for $\sim5-10$ light-crossing times to ensure that the distribution functions of all particle species have reached a constant value.} model that describes adequately the multi-wavelength data for each epoch. More specifically, we report those parameters for which the model curve passes through most of the instrument-specific SED bands, i.e., we perform a ``fit by eye", as commonly done in blazar modeling studies \citep[e.g.,][]{Tavecchio2010, Abdo2011, Boettcher:2013wxa, Petropoulou:2015upa, Petropoulou:2016ujj, Gao:2016uld}. 

After having determined the values of those parameters that were allowed to vary in order to explain the multi-epoch SEDs in terms of synchrotron and inverse Compton emissions from primary electrons, we can proceed with the derivation of the maximal proton luminosity $L^{\prime (\max)}_{p}$. This, in turn,  translates into an upper limit on the 
blazar's neutrino flux, $F_{\nu+\bar{\nu}}^{(\max)}$, for each epoch. To do so, we require that any proton-induced emission does not overshoot the X-ray and/or gamma-ray data \citep{Keivani2018, Gao2019}. 
Rough estimates of $L^{\prime (\max)}_{p}$ and $F_{\nu+\bar{\nu}}^{(\max)}$ can be obtained from the X-ray flux using analytical arguments \citep{Murase2018, Padovani:2019}. However, an accurate determination of these upper limits requires knowledge of the spectrum emitted by all secondaries produced in photo-pair, photomeson, and photon-photon pair production processes. As the production rates and energy spectra of the produced secondaries are sensitive to the conditions in the emitting region (e.g., density and spectrum of photons), it is important to benchmark these parameters with the leptonic SED modeling.

\begin{figure*}
\centering
\includegraphics[width=0.48\textwidth]{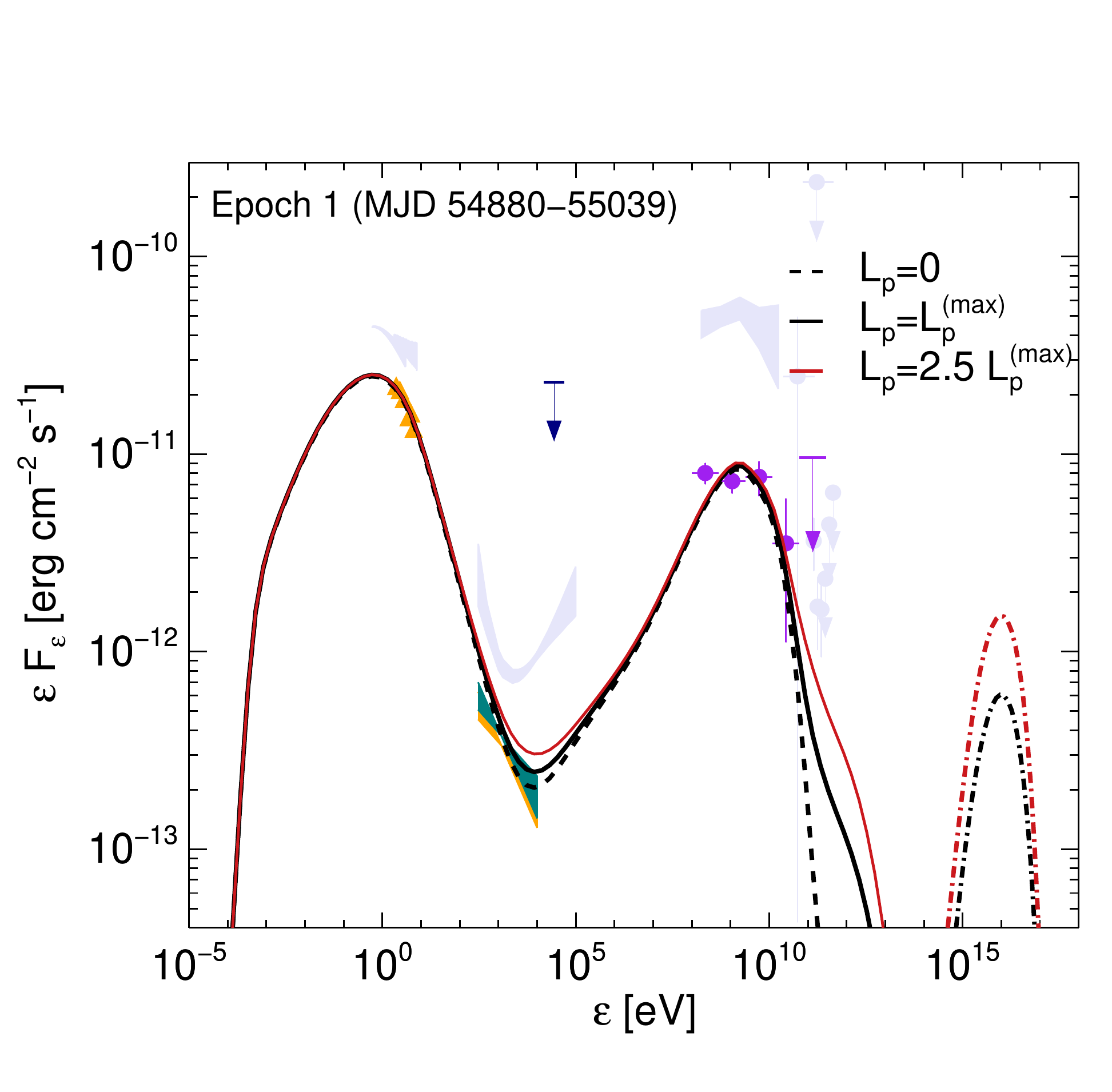}
\includegraphics[width=0.48\textwidth]{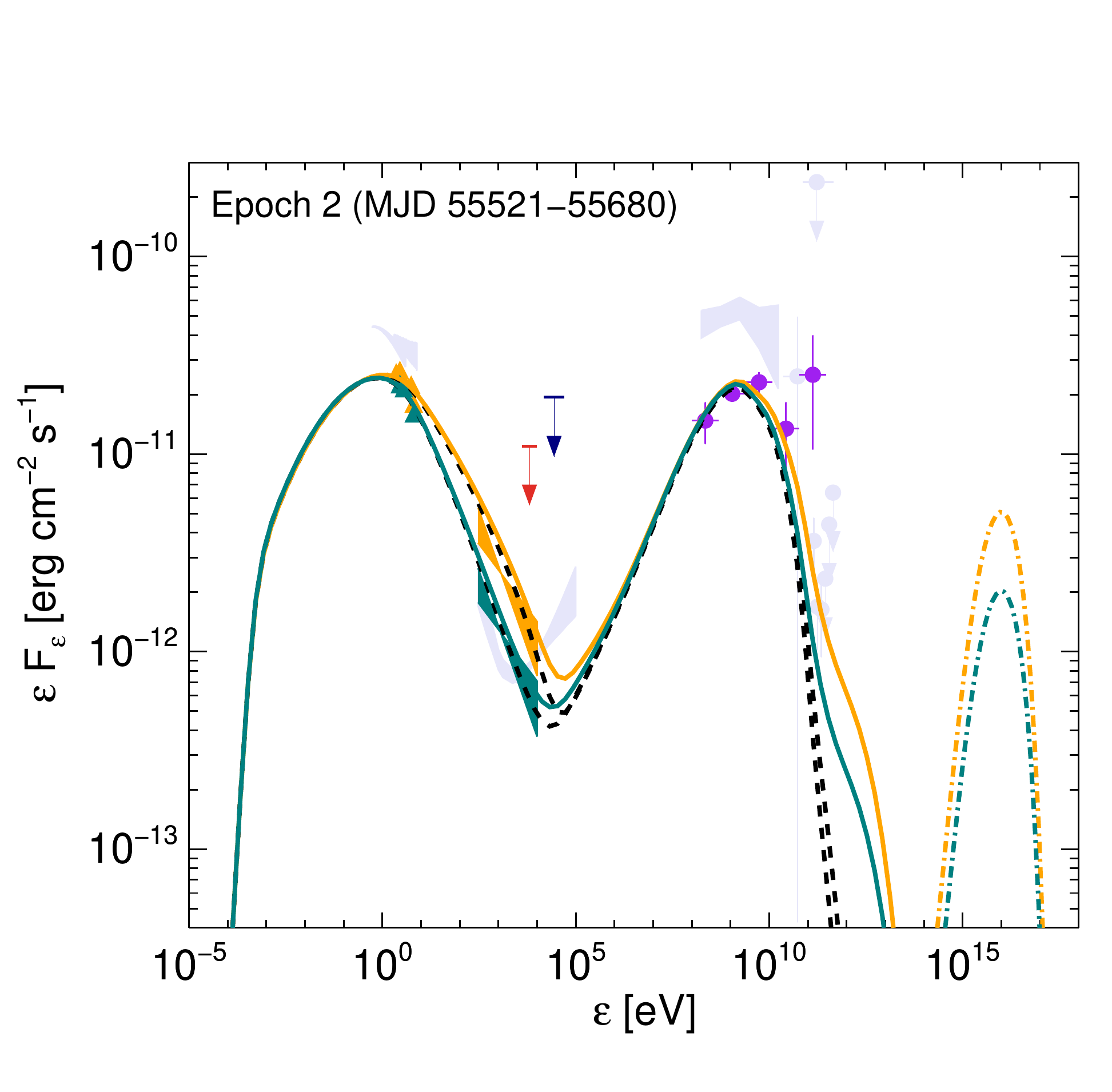}
\includegraphics[width=0.48\textwidth]{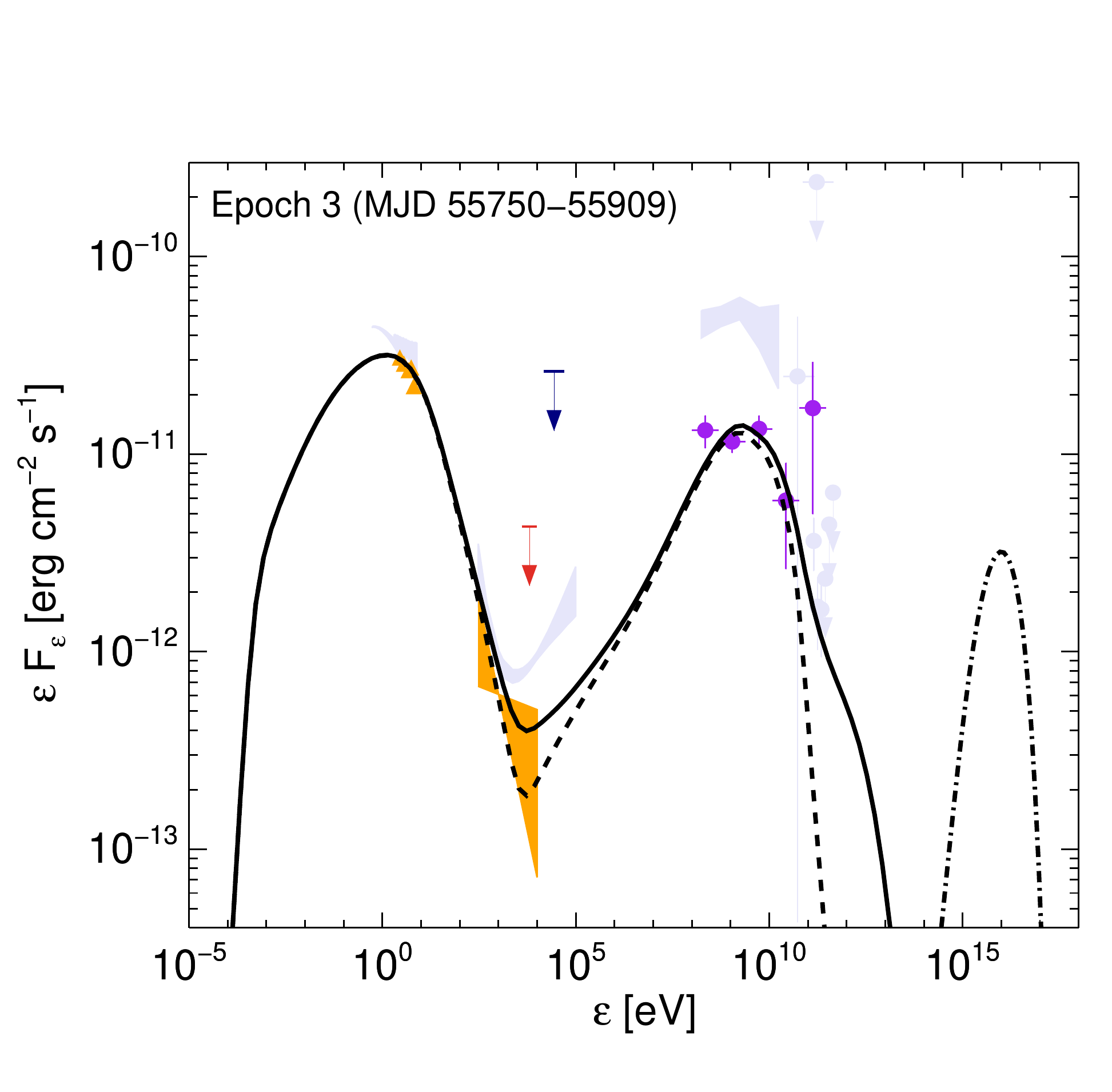}
\includegraphics[width=0.48\textwidth]{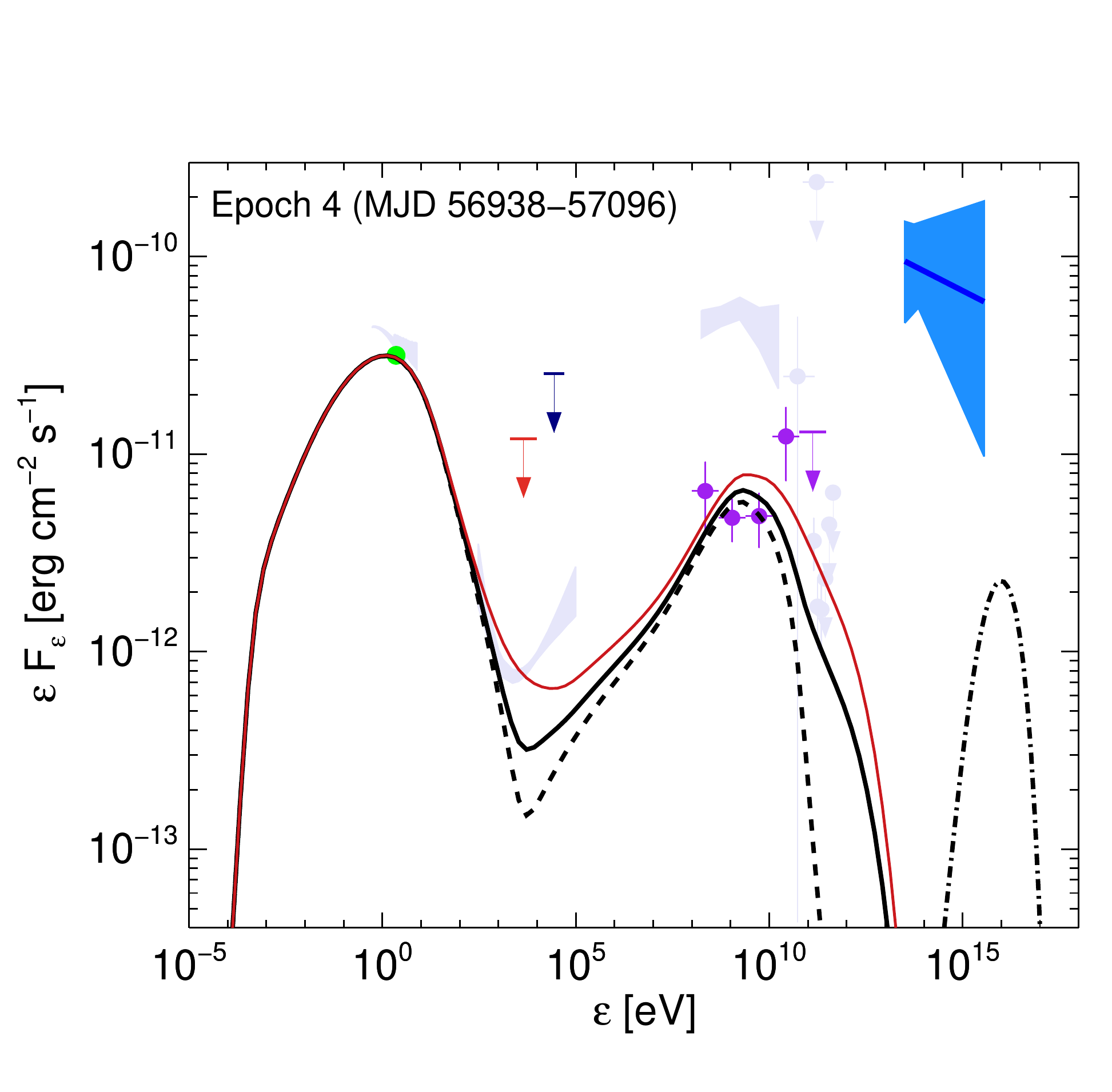}
\caption{Multi-epoch average SEDs of \txs \ built with data from \fermi-LAT (filled purple circles), \swift-XRT (colored bow ties), \swift-UVOT (filled colored triangles) and ASAS-SN (filled green circle). The $3\sigma$ upper limits from \maxi \, and \swift-BAT are shown as red and blue arrows, respectively. Data from the six-month long flare of \txs\, in 2017 \citep[filled pale lavender symbols and bow ties, adopted from][]{Keivani2018,Abeysekara2018} are overplotted for comparison. In each panel, we show the photon spectra computed for a pure leptonic model (dashed black lines) and a hybrid leptonic model with the proton luminosity set to its maximum-allowed value $L_p^{(\max)}$ (solid black lines). The maximal all-flavor neutrino fluxes  from the hybrid model (dashed-dotted lines) are also shown. For illustration purposes, we also show the results obtained for $2.5 L_p^{(\max)}$ during epochs 1 and 4 (red lines).
For epoch 2, we show two models to account for the X-ray flux variability of the \swift-XRT observations. The blue-colored bow tie (epoch 4) shows the best-fit  all-flavor spectrum (with its 68\% uncertainty region) obtained by IceCube \citep[adopted from Figure 3 of][]{Aartsen2018blazar2}. Photon attenuation at $\varepsilon_{\gamma}\gtrsim3\times10^{11}$\,eV due to interactions with the extragalactic background light is not included here.}
\label{fig:SEDs}
\end{figure*}

Our results are summarized in Table~\ref{tab:fit} and Figure~\ref{fig:SEDs}. The SEDs of a pure leptonic model are indicated with dashed lines, while solid lines show the SEDs of the hybrid leptonic model obtained for the maximal proton luminosity $L^{\prime (\max)}_p$. The associated all-flavor neutrino flux is also shown in each panel with a dashed-dotted line. The multi-epoch SEDs can be described satisfactorily\footnote{Note that the model spectrum is too steep to account for the gamma-ray flux above $\sim 10$~GeV for epochs 2 to 4.} by essentially varying only several model parameters. The energy density of the external photon field $u^\prime_{\rm ext}$ and the  electron injection luminosity $L_{e}^\prime$ are varied to mainly account for the variable gamma-ray and optical/UV fluxes. The former parameter changes no more than a factor of $5$ compared to its value used in modeling the 2017 flare, while $L_{e}^\prime$ changes at most by a factor of $2.5$ across all epochs (see Table~\ref{tab:fit}). In addition, only small changes in the properties of the electron energy spectrum at injection (i.e., $\gamma_{e, b}^{\prime}$, $s_{e,2}$, $\gamma_{e, \max}^{\prime}$) are needed to account for the spectral variability in the X-ray band. Thus, despite the apparent large differences in the observed SEDs between the 2017 flare and the epochs considered here, only small changes of the parameters are needed (with respect to their baseline values).

In all epochs, we find that the \fermi-LAT observations are satisfactorily explained by the inverse Compton scattering of external photons by the electrons in the blazar jet (external Compton; EC), while the contribution of the Compton-scattered synchrotron emission (synchrotron self-Compton; SSC) to the total spectrum is sub-dominant (not explicitly shown in Figure~\ref{fig:SEDs}). The relation among the various components of the primarily leptonic SEDs can be understood by comparing the relevant energy densities (i.e., of the magnetic field and of the seed photons for Compton scattering). For all epochs considered here, we find that the co-moving synchrotron energy density is $u^\prime_{\rm syn} \simeq (1.5-2)\times10^{-4}$~erg cm$^{-3}$, while $u^\prime_{\rm ext}\simeq (0.7-2.9)\times10^{-2}$~erg cm$^{-3}$ (see Table~\ref{tab:fit}). Given that the magnetic energy density is $u^\prime_{\rm B}\simeq 6\times10^{-3}$~erg cm$^{-3}$, we find that $u^\prime_{\rm ext} \gtrsim u^\prime_{\rm B} \gg u^\prime_{\rm syn}$, for all epochs including the 2017 flare. This relation suggests that the synchrotron and EC components should have similar peak fluxes (modulo differences because of the spectral broadness of each component), while the SSC peak flux should be order of magnitude less than the synchrotron peak flux. Similar results were reported by \citet{Keivani2018} for the 2017 flare, where the rising part of the SSC component could be, however, constrained by the \emph{NuSTAR} data. 

The strongest upper limits on the neutrino flux are derived for epochs where X-ray measurements (with small uncertainties) are available, in agreement with previous studies~\citep{Keivani2018, Murase2018}. Unfortunately, during the period of the reported neutrino flare by IceCube, there are no X-ray measurements available. As a result, our hybrid leptonic model cannot be constrained as well as for the other epochs, and the derived upper limit on the neutrino flux should be considered as an optimistic prediction.
 
\begin{figure*}
\centering 
\includegraphics[width=0.47\textwidth]{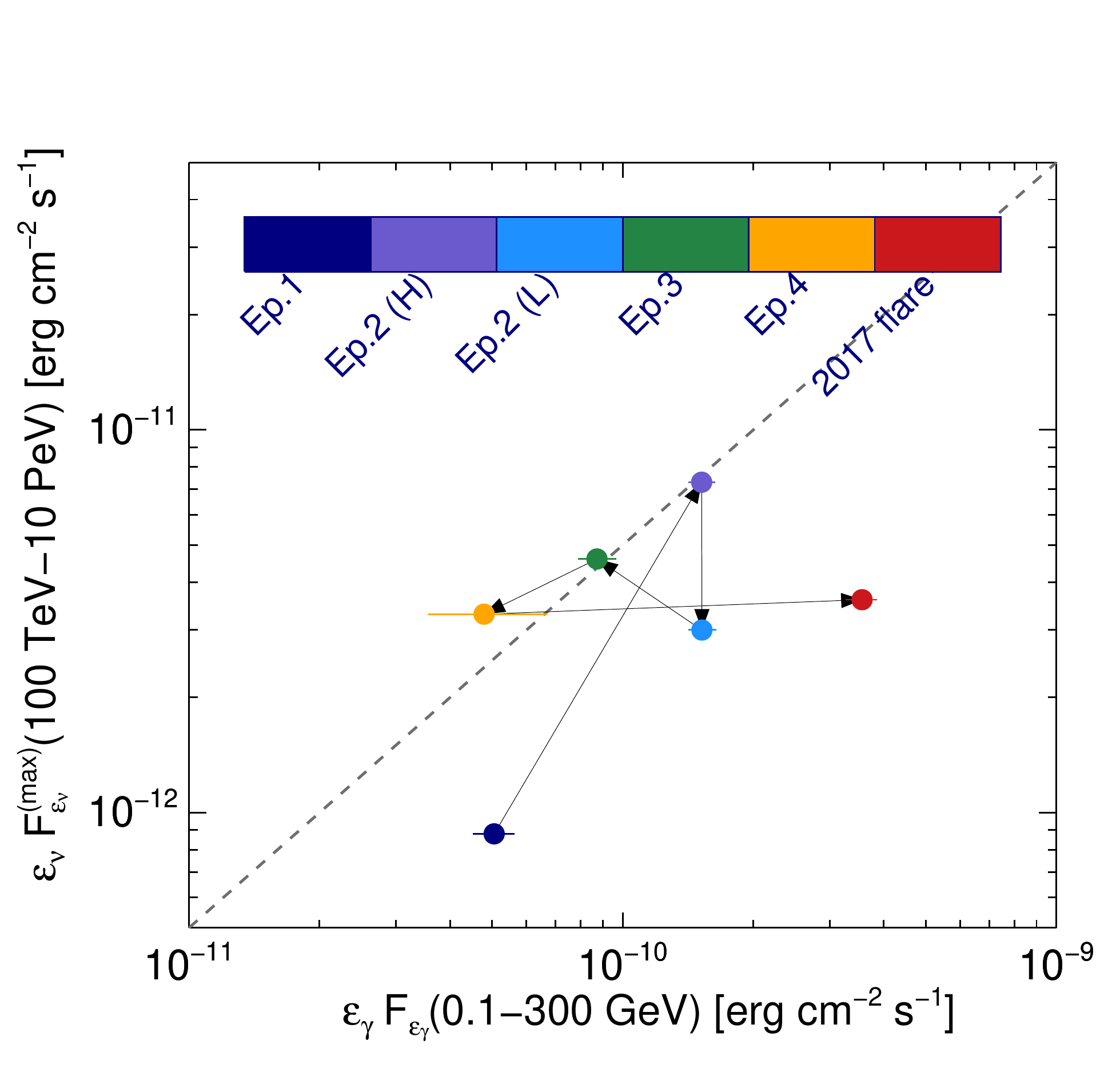}
\includegraphics[width=0.47\textwidth]{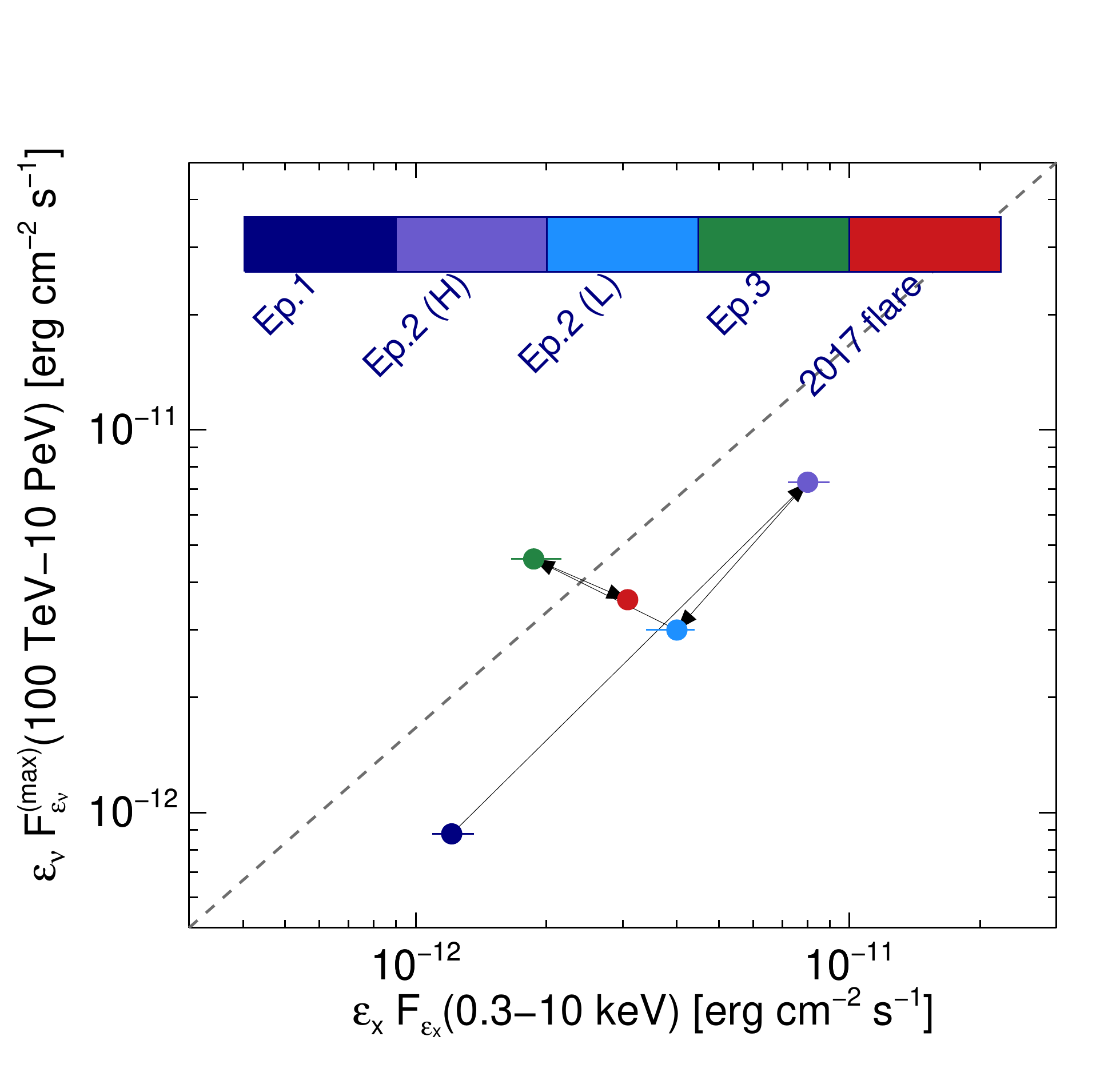}
\caption{Maximal model-predicted all-flavor neutrino flux (100 TeV -- 10 PeV) versus the \fermi-LAT ($0.1-300$ GeV) gamma-ray flux (left panel) and the \swift-XRT ($0.3-10$~keV) X-ray flux (right panel) for different epochs in chronological order (see inset color bar). The dashed gray line has a slope of one and is plotted to guide the eye. The integrated gamma-ray fluxes are computed using the best-fit power-law model reported in Table~\ref{tab:Fermi} and the X-ray fluxes for epochs 1-3 are taken from Table~\ref{tab:XRT}. The fluxes for the 2017 flare are adopted from \citet{Keivani2018}.} 
\label{fig:flux-flux}
\end{figure*}
 
Figure~\ref{fig:flux-flux} shows the maximal all-flavor neutrino flux (100 TeV--10 PeV) derived from the multi-epoch SED modeling against the best-fit gamma-ray and X-ray fluxes in the 0.1--300 GeV and 0.3--10 keV energy bands, respectively (see Tables~\ref{tab:Fermi} and \ref{tab:XRT}). Arrows are overplotted to indicate more clearly the temporal evolution of the fluxes across epochs. Similar results as those shown in the left panel of the figure are obtained if we plot the gamma-ray flux of the model computed in the energy range of proposed MeV detectors, such as \emph{AMEGO} \citep{AMEGO_2019} and \emph{e-ASTROGAM} \citep{eASTROGAM_2018}. In the hybrid leptonic scenario considered here, we find that the maximal neutrino flux follows more closely changes in the soft X-ray flux  probed by \swift-XRT (i.e., up to 10 keV) instead of the gamma-ray flux in the \fermi-LAT band\footnote{A stronger correlation with gamma-rays is expected in leptohadronic scenarios for the blazar gamma-ray emission \citep[e.g.,][]{Murase:2014foa,Petropoulou:2016ujj}.}, as previously argued by \citet{Murase2018} with analytical considerations. In this context, the 2017 major gamma-ray flare does not seem to be a special period in terms of neutrino production. On the contrary, the highest maximal neutrino flux is found for epoch 2, where the source was in an X-ray bright state.

\section{Implications for IceCube Observations}\label{sec:long} 
We estimate the number of through-going muons induced by muon neutrinos, expected from \txs \, over the IceCube lifetime and discuss the implications of our results for the 2014-2015 neutrino flare. 

\subsection{Long-term neutrino emission of \txs}
Assuming that the model-predicted maximal neutrino flux for each epoch is representative for the long-term neutrino emission of \txs \, we calculate the number of through-going muon tracks~\citep{Laha:2013lka,murasewaxman16}, integrated above a muon energy of 100~GeV. As we are interested in the numbers that can be obtained by standard point-source analyses, we do not use the effective areas for real-time alerts\footnote{For the neutrino flux of the 2017 multi-messenger flare, \cite{Keivani2018} obtained $\dot{\mathcal{N}}_{\nu_\mu+\bar{\nu}_\mu}\sim0.02~{\rm yr}^{-1}$ for the real-time effective area and $\dot{\mathcal{N}}_{\nu_\mu+\bar{\nu}_\mu}\sim0.1~{\rm yr}^{-1}$ for the point-source effective area, respectively.}. Our results are summarized in Table~\ref{tab:neutrinos}. 

If we take epochs 1 and 2 (\swift-XRT high state) as the most pessimistic and optimistic periods, respectively, for the neutrino emission of \txs, then we obtain the following range for the number of muon neutrinos expected in ten years of IceCube observations:   $\mathcal{N}_{\nu_\mu+\bar{\nu}_\mu}\sim(0.4-2)~(\Delta T/10~{\rm yr})$. Our model predictions for the neutrino emission during the 2014-2015 and 2017 flares lie in that range. The detection of one high-energy neutrino event associated with the 2017 gamma-ray flare is consistent with an upper fluctuation from the average event rate. This is not surprising given that no correlation of muon neutrinos with blazars has been found in stacking analyses~\citep{ic17_blazars,Neronov:2016ksj,Hooper:2018wyk,Oikonomou:2019djc,Yuan:2019ucv}. Our single-zone prediction for epoch 4, however, significantly underestimates the observed number of neutrinos ($13\pm5$) in 2014-2015. We discuss this result in more detail in the next subsection.

The predicted number $\sim 0.4-2$ of muon neutrinos in ten years should be considered as a promising signal from the blazar zone of \txs, as long as there is a variable external photon field (on month-long timescales) and the jet's power in relativistic protons is $L_{p, j}^{(\max)} \approx (0.9-6) \times 10^{47}$~erg s$^{-1}$ (see Table~\ref{tab:fit}). 
Non-detections in twenty years of IceCube observations will exclude the most optimistic case and constrain the proton power of the jet \citep[for constraints on other individual blazars, see][]{Aartsen2017, Aartsen2018}.
These predictions can be further tested over a shorter period of time with the next-generation neutrino telescope IceCube-Gen2, which is expected to have $\sim5$ times larger effective area than IceCube.

\begin{deluxetable}{lccc}
\tablecaption{Upper limits on the 100 TeV -- 10 PeV all-flavor neutrino flux and muon neutrino rate for muons above 30~TeV.\label{tab:neutrinos}}
\tablewidth{0pt}
\tablehead{\colhead{Epoch} & \colhead{$F_{\nu+\bar{\nu}}^{(\rm max)}$ [erg cm$^{-2}$ s$^{-1}$]} & \colhead{$\dot{\mathcal{N}}_{\nu_\mu+\bar{\nu}_\mu}$ [yr$^{-1}$]}
}  
\startdata
1  & $8.8\times10^{-13}$ & 0.04 \\ 
2\tablenotemark{\dag}  & $7.3\times10^{-12}$  & 0.2 \\
2\tablenotemark{\ddag} & $3.0\times10^{-12}$  &  0.1 \\
3  & $4.6\times10^{-12}$  &  0.2 \\
4  & $3.3\times10^{-12}$  & 0.1 \\ 
2017& $3.6\times10^{-12}$ & 0.1 \\
\enddata 
\tablecomments{We also list the value for the LMBB2b model of  \citet{Keivani2018} for the 2017 flare of \txs. The atmospheric background muon neutrino rate in the 100 TeV -- 10 PeV energy range is $\dot{\mathcal{N}}_{\nu_\mu+\bar{\nu}_\mu}^{\rm atm}\sim0.01$~yr$^{-1}$ for an angular resolution of $0.5$~deg.}
\tablenotetext{\dag}{\swift-XRT high state.} \tablenotetext{\ddag}{\swift-XRT low state.}
\end{deluxetable}

We next discuss a few caveats that should be kept in mind when interpreting our predictions for the long-term neutrino emission of \txs. 
\begin{enumerate}
    \item The predictions rely on the assumption that the maximal neutrino flux obtained for each epoch is representative of the long-term neutrino emission of the source. Ideally, one should find a scaling relation between the maximal neutrino flux and the photon flux in some energy band with continuous temporal coverage, and then use the long-term light curve to compute the predicted number of muon neutrinos \citep[e.g.,][]{Petropoulou:2016ujj}. Although the 0.1--300 GeV energy band of \fermi \, is ideal for this purpose, we cannot establish a robust relation between $F_{\nu+\bar{\nu}}^{(\max)}$ and $F_{\gamma}$, as shown in Figure~\ref{fig:flux-flux} (left panel). On the contrary, we find that the X-ray flux is a better probe of the maximal neutrino flux within our model, with $F_{\nu+\bar{\nu}}^{(\max)} \propto F_{X}$ (right panel of Figure~\ref{fig:flux-flux}). This is partly because the SED has a valley in the X-ray range, which is the most important for constraining hadronic components. The X-ray coverage of the source before the 2017 flare is very sparse (see Figure~\ref{fig:LC}), thus preventing a more sophisticated analysis than the one presented here. 
    \item We cannot exclude the possibility that the physical properties of the jet change drastically in-between the four epochs we chose for our analysis.  Such changes in the jet parameters could happen in highly variable blazars~\citep[e.g.,][]{Raiteri:2013MNRAS, Ahnen2017}.
    This limitation stems from the lack of quasi-simultaneous multi-wavelength data for long time windows and highlights the need for X-ray monitoring of blazars.
    \item The SEDs we constructed are not contemporaneous. More specifically, the X-ray spectra are computed from individual \swift-XRT observations of duration of few ks each, while the gamma-ray spectrum is averaged over the whole epoch of interest ($\sim 0.5$~yr). In this regard, the \swift-XRT observations are instantaneous compared to the selected time window. So, when we translate the maximal neutrino flux, which is mainly set by the X-ray flux, into an expected number of events and use $\Delta T = 0.5$~yr as the typical duration, we may overestimate the number of neutrinos. The X-ray flux variability within epoch 2, for example, can lead to an overestimation of the neutrino number by a factor of $\sim2$.
\end{enumerate}

\begin{figure}
\centering
\includegraphics[width=0.47\textwidth]{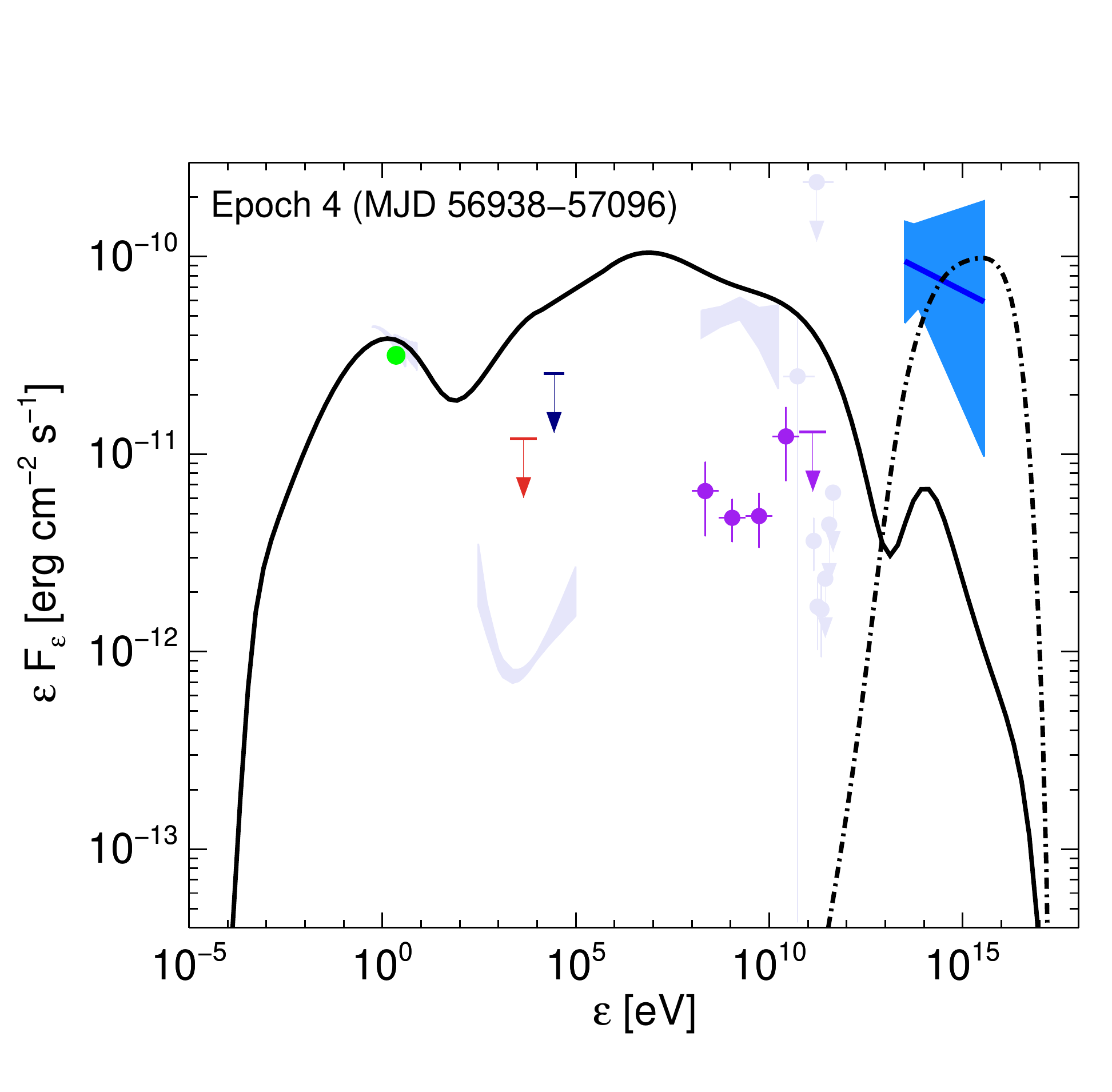}
\caption{Same as in Figure \ref{fig:SEDs}, but for a case where the model-predicted neutrino flux is compatible with the IceCube flux of epoch 4. Here, we assumed $T_{\rm ext}^\prime=2\times10^7$~K (or, equivalently, $\epsilon^\prime_{\rm ext}\simeq 5$~keV) and  $L^\prime_p=1.7\times10^{48}$~erg s$^{-1}$. All other parameters are the same as those listed in Table~\ref{tab:fit} for epoch 4.   
}
\label{fig:excess}
\end{figure}

\subsection{Implications for the 2014-2015 neutrino flare}
Here, we focus to the implications of our model for the 2014-2015 neutrino flare.  
As an illustrative example, we show in Figure \ref{fig:excess} a case where the model-predicted neutrino flux is compatible with the IceCube flux of epoch 4. The parameters are the same as those listed in Table~\ref{tab:fit}, except for the characteristic external photon energy (temperature) and the proton luminosity, which now read $\epsilon^\prime_{\rm ext}\simeq 5$~keV ($T_{\rm ext}^\prime=2\times10^7$~K) and $L^\prime_p=1.7\times10^{48}$~erg s$^{-1}$, respectively. 
For the adopted parameters, the electromagnetic emission of the secondaries produced via photohadronic interactions and photon-photon pair production  reaches a flux of $\sim(3-10)\times10^{-11}~{\rm erg}~{\rm cm}^{-2}~{\rm s}^{-1}$, which confirms the analytical results of \citet{Murase2018}. Such high X-ray and gamma-ray fluxes clearly overshoot the \maxi, \swift-BAT upper limits by a factor of $\sim2-3$ and the \fermi-LAT data by a factor of $\sim 10$, respectively. In addition, this case is unlikely in astrophysical view, for it requires a highly super-Eddington proton power to account for the low photomeson production efficiency.

Given the unprecedented neutrino flux measured by IceCube in 2014--2015, one could still argue that the conditions in the blazar zone were significantly different compared to other epochs. We therefore explored this possibility, by performing a wide scan of the parameter space for one-zone models. Our methodology and results are presented in Appendix~\ref{sec:appendix-1}. We found no parameter set for the blazar zone that can simultaneously explain the neutrino flare and be compatible with the electromagnetic constraints. Moreover, all cases require a highly super-Eddington jet power, namely $(10^2-10^3)\, L_{\rm Edd}$, where $L_{\rm Edd}\simeq 1.3\times 10^{47} \, (M/10^9 M_{\odot})$~erg s$^{-1}$ is the Eddington luminosity of a black hole with mass $M$. The necessary proton power could be reduced to Eddington levels, if the energy density of the external photon field (in the blazar zone) was two or three orders of magnitude higher than all other epochs~\citep[see also][]{Reimer2019}. 

We therefore conclude that the high neutrino flux of epoch 4 cannot be explained concurrently with the electromagnetic data, if both emissions originate from the same region, in agreement with previous studies ~\citep{Murase2018,Reimer2019,Rodrigues2019}.

\section{Discussion}\label{sec:discussion}
\subsection{Remarks on the maximal neutrino flux and proton luminosity}
 We have constrained the maximal neutrino flux ($F_{\nu+\bar{\nu}}^{(\max)}$) and the required proton luminosity ($L_p^{(\max)}$), assuming that the low-energy hump in the SED is attributed to synchrotron emission from primary electrons. This assumption is plausible and widely accepted. Indeed, the optical-to-soft X-ray data can be fitted with a single power law,  especially evident in epoch 2 and in the 2017 flare~\citep{Keivani2018}. It is therefore unlikely that proton-initiated cascades \citep[with usually broad curved energy spectra, see e.g.,][]{petromast15, Petropoulou:2015upa,Keivani2018,Gao2019} overtake the leptonic emission in the X-ray range. As long as the low-energy hump of the SED is predominantly explained by primary electrons, the results on $F_{\nu+\bar{\nu}}^{(\max)}$ and $L_p^{(\max)}$ are rather robust against changes in source parameters, because the cascade emission is broad and represents the reprocessing of particle energy injected by cosmic rays~\citep[cf.][for the 2017 multi-messenger flare]{Murase2018,Keivani2018,Gao2019}.

\begin{figure}
\centering   \includegraphics[width=0.47\textwidth, trim={0, 0, 0, 40}]{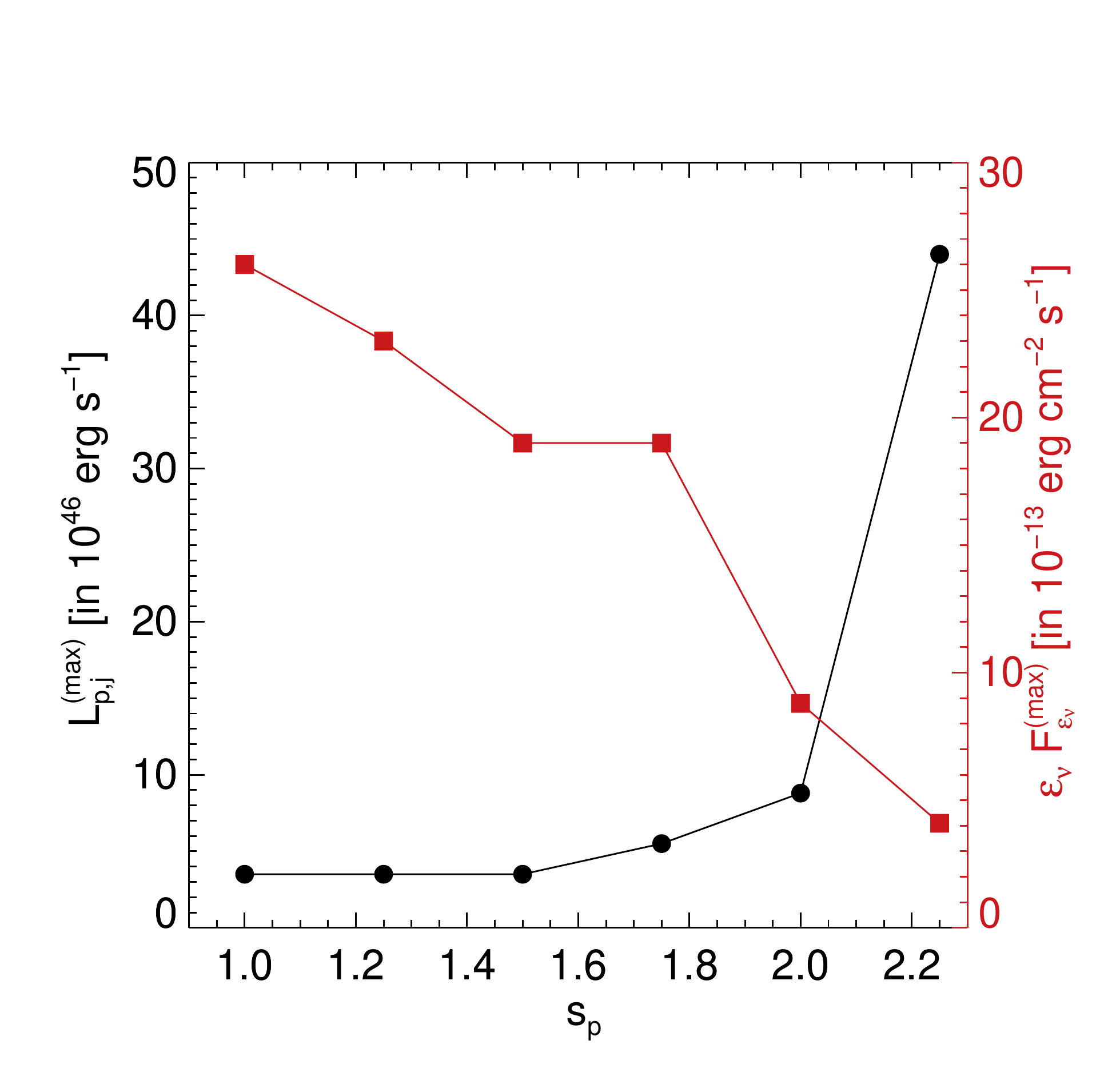}
\caption{Maximal jet power in protons (black circles) and maximal all-flavor neutrino flux in the 100 TeV -- 10 PeV energy range (red squares) for epoch 1 as a function of the power-law proton index $s_p$.}
\label{fig:proton-param}
\end{figure}

Although a detailed investigation of all proton and electron parameters is definitely beyond the scope of this paper, we discuss the effect of two (unconstrained by the data) parameters of the proton distribution, namely the power-law index $s_p$ and the minimum Lorentz factor $\gamma^\prime_{p, \min}$, on $L_{p,j}^{(\max)}$ and $F_{\nu+\bar{\nu}}^{(\max)}$ (for the effects of $\gamma_{p,\max}^\prime$ we refer the reader to \citealt{Keivani2018}). For the purposes of this discussion, we choose epoch 1 where  the uncertainties in the X-ray spectrum are small, but similar trends are expected for the other epochs as well. Our results for $s_p$ are summarized in Figure~\ref{fig:proton-param}. Harder proton energy spectra (i.e., $s_p<2$) tend to decrease the maximal proton luminosity, but no more than a factor of $\sim 2.5$. In contrast, $L_{p,j}^{(\max)}$ increases rapidly with $s_p>2$, as most of the energy is carried by low energy protons that do not participate in the photohadronic interactions. For a fixed target photon field, the flux of secondaries produced in the photomeson production process should increase with decreasing $s_p$, since more power is carried by protons with higher energies relevant for neutrino production \citep[see also Figure 12 in][]{DMPR12}. Indeed, as shown in Figure~\ref{fig:proton-param}, the maximal neutrino flux increases as the proton energy spectra become harder. In the optimistic scenario with $s_p=1$, the maximal neutrino flux can be $\sim 3$ times higher than the value reported in Table~\ref{tab:neutrinos}.  
Note that the required luminosity is not sensitive to $\gamma^\prime_{p, \min}$ for $s_p<2$. For $s_p=2$, we find that $L_{p,j}^{(\max)}$ decreases at most by a factor of $\sim 2.5$, while $F_{\nu+\bar{\nu}}^{(\max)}$ increases by the same factor when $\gamma_{p,\min}$ increases by six orders of magnitude. Similar trends are found for $s_p>2$, but the quantitative changes are larger. Henceforth, we report the maximal neutrino fluxes derived for the default choice of $s_p=2$ and $\gamma_{p,\min}=1$.

\subsection{Remarks on the baryon loading factor}
\begin{figure}
\centering
\includegraphics[width=0.47\textwidth]{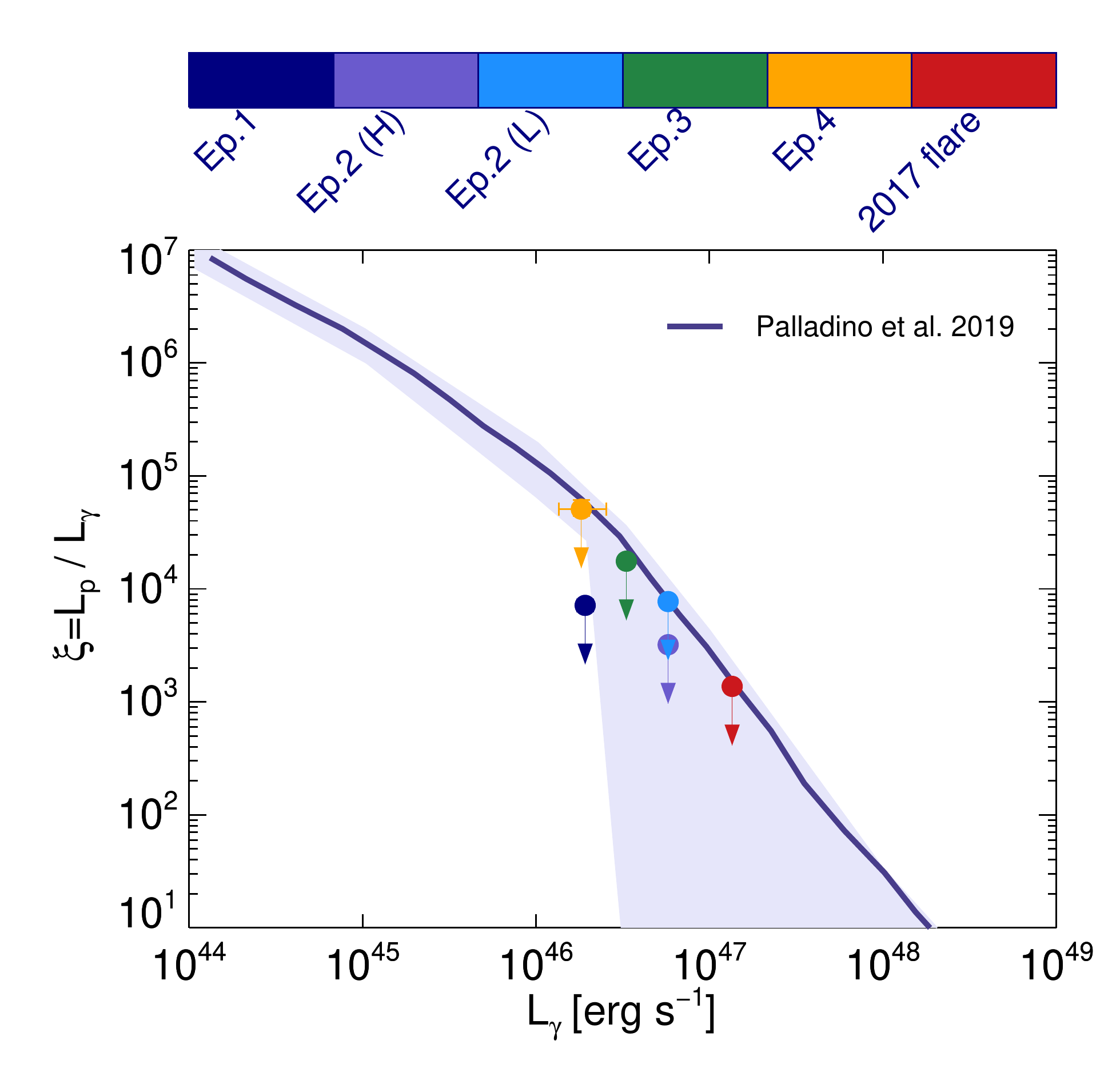}
\caption{Maximal baryon loading factor $\xi$ of \txs \, (filled symbols with arrows) as a function of the \fermi-LAT (0.1--300 GeV) gamma-ray luminosity for different epochs (see colorbar). For comparison, we show the baryon loading factor (solid blue line) with its uncertainty (shaded region) obtained from a model for the diffuse astrophysical neutrino flux at energies $\gtrsim 1$~PeV from blazars \citep[see scenario 3 in][]{palladino_2019}.}
\label{fig:baryon}
\end{figure}

From our analysis, we can also determine the maximal baryon loading factor, defined as $\xi^{(\max)}\equiv L^{(\max)}_{p}/L_{\gamma}$, where $L_{\gamma}$ is the gamma-ray luminosity in the 0.1--300 GeV energy band. Our results for the different epochs considered in this study as well as for the 2017 flare \citep{Keivani2018} are presented in Figure~\ref{fig:baryon} (filled symbols with arrows). 
The upper limits are much larger than the values required for all blazars to explain the flux of ultra-high-energy cosmic rays, $\xi\sim3-100$~\citep{Murase:2014foa}.

 Nonetheless, our results can constrain some extreme models. 
For reference, we also show the baryon loading factor and its uncertainty (solid blue line and shaded region) invoked to explain the diffuse astrophysical neutrino flux at energies $\gtrsim1$~PeV with blazars~\citep[for details, see][]{palladino_2019}. Although there is no physically motivated scenario to predict a negative correlation between the baryon loading factor and gamma-ray luminosity, our results demonstrate that multi-epoch modeling of even a single source at different gamma-ray luminosity levels can be a powerful method for constraining models of diffuse neutrino emission from blazars. Similar studies of individual sources, spanning a wide range of gamma-ray luminosities, are strongly encouraged.

\begin{deluxetable}{lcc}
\tablecaption{Upper limit on the ratio $Y_{\nu \gamma} \equiv L_{\nu+\bar{\nu}}/L_{\gamma}$ obtained for different epochs, and gamma-ray luminosity in the $0.1-300$ GeV energy range, $L_{\gamma}$. \label{tab:ratio}}
\tablehead{\colhead{Epoch} & \colhead{$L_{\gamma}$ [erg s$^{-1}$]} & \colhead{$Y_{\nu\gamma}^{(\max)}$}
}
\startdata
1  & $1.9^{+0.2}_{-0.2}\times10^{46}$ & 0.017\\
2\tablenotemark{\dag}  & $5.8^{+0.4}_{-0.4}\times10^{46}$ & 0.048\\
2\tablenotemark{\ddag} & $5.8^{+0.4}_{-0.4}\times10^{46}$ & 0.020 \\
3  & $3.3^{+0.4}_{-0.3}\times10^{46}$  & 0.053 \\
4  & $1.8^{+0.7}_{-0.5}\times10^{46}$ & 0.069 \\
2017& $1.3^{+0.1}_{-0.1}\times10^{47}$ & 0.010\\
\enddata 
\tablecomments{$L_{\gamma}$ is computed using the best-fit power-law model reported in Table~\ref{tab:Fermi}, except for the 2017 flare, for which we adopted the value from \citet{Keivani2018}.}
\tablenotetext{\dag}{\swift-XRT high state.} \tablenotetext{\ddag}{\swift-XRT low state.}
\end{deluxetable} 

A blazar's total neutrino luminosity is commonly parametrized as $L_{\nu+\bar{\nu}}=Y_{\nu\gamma} \, L_{\gamma}$, where $Y_{\nu\gamma}$ encodes information about the baryon loading and neutrino production efficiency of the source~\citep{Petropoulou:2015upa, Padovani:2015mba, palladino_2019}\footnote{Roughly speaking, one sees $Y_{\nu\gamma}\sim(3/8)f_{p\gamma}\xi$, where $f_{p\gamma}$ is the efficiency of the photomeson production.}. The multi-epoch upper limits on the ratio $Y_{\nu \gamma}$ for \txs \, are summarized in Table~\ref{tab:ratio}. We find $Y_{\nu \gamma} \sim 0.01-0.07 \ll 1$. These values are suggestive of a leptonic origin for the gamma-ray emission and are consistent with our initial assumption of a leptonic SED \citep[see also][]{Murase:2014foa,Petropoulou:2015upa}.  
\citet{Padovani:2015mba} computed the contribution of BL Lac sources to the diffuse neutrino flux, assuming a constant ratio $Y_{\nu \gamma}=0.8$ for all blazars, which has been constrained by IceCube upper limits on the diffuse neutrino flux at extremely high energies (i.e., $\gtrsim 1$~PeV)as $Y_{\nu \gamma} \lesssim 0.1$~\citep{Aartsen:2016prl}. Thus, if the hybrid leptonic model was to be applied to the whole BL Lac population, assuming a universal ratio $Y_{\nu \gamma}$ in the range $0.01-0.07$ as found for \txs, then the model's predictions would be consistent with the latest upper limits from IceCube~\citep{Aartsen:2018PhRvD}. However, this implies that the contribution to the diffuse neutrino flux must be sub-dominant.

\subsection{Possible origins of the external radiation field}
In general, the location of the gamma-ray emitting region and the origin of external seed photons for inverse Compton scattering in luminous flat-spectrum radio quasars (FSRQs) has been under debate \citep[for a recent review, see][]{Dermer2016}. Identifying the source of seed photons is also critical for understanding the electromagnetic emission and neutrino production of \txs.\,  
Recently, \citet{Padovani:2019} pointed out that \txs \ is in fact a masquerading BL Lac object, namely a blazar with a broad line region (BLR) whose radiation is, however, swamped by the non-thermal jet emission \citep{Blandford1978, Georganopoulos1998}. Using various diagnostics, \citet{Padovani:2019} showed that the black hole of \txs \, is accreting matter from a disk with luminosity $L_{\rm AD} \approx 8\times10^{44}$~erg s$^{-1}$. The estimated BLR luminosity and radius are $L_{\rm BLR}\approx 5\times 10^{43}$~erg s$^{-1}$ and $R_{\rm BLR}\approx7\times10^{16}$~cm, respectively. We can relate the radius of the emission region $R^\prime$ to the distance of the blazar zone from the black hole ($r_{\rm BZ}$), by assuming  that the blob covers the whole cross-sectional area of a conical jet with opening angle $\theta_j\approx 1/\Gamma$, namely $r_{\rm BZ}\approx R^\prime/ \theta_j \approx 0.3 \, (R^\prime/10^{17} \ {\rm cm})(\Gamma/10) \ {\rm pc}$. Given that $r_{\rm BZ} \gg R_{\rm BLR}$, the BLR radiation would appear de-boosted (dilute) in the jet's comoving frame \citep[e.g.,][]{Sikora2009} with $u^\prime_{\rm BLR} \approx L_{\rm BLR}/4 \pi R_{\rm BLR}^2 c \Gamma^2 \ll u^\prime_{\rm ext}$. Thus, the BLR could not account for the energy densities needed to explain the gamma-ray emission within our model.   

Other possible origins for the putative external photon field are the disk-scattered emission and the radiation from an outer layer of the jet \citep{Murase:2014foa,Dermer:2014vaa,Tavecchio2015}. If there is a scattering region with Thomson optical depth $\tau_T$ at parsec-scale distances,  then the energy density of the scattered emission is $u_{\rm sc}\lesssim 1.6 \times 10^{-4}(\tau_T/0.1) (L_{\rm AD}/8\times10^{44}\ {\rm erg \ s^{-1}})(r_{\rm BZ}/10^{18}~{\rm cm})^{-2}$~erg cm$^{-3}$. Given that $u_{\rm ext}\approx u_{\rm ext}^\prime/\Gamma^2\simeq (6.6-33)\times10^{-5} (\Gamma/10)^{-2}$~erg cm$^{-3}$ (see Table~\ref{tab:fit}), the scattered-disk radiation is an energetically viable scenario. However, the origin of the scattering material is less clear.

Alternatively, we can assume that \txs \, has a structured jet composed of a fast spine with $\Gamma_s \gg 1$ and a slower outer layer with $\Gamma_l < \Gamma_s$.  The Lorentz factor of their relative motion is then $\Gamma_{\rm rel} \approx \Gamma_s / (2\Gamma_l) \approx 2.5 (\Gamma_s/10) (2/\Gamma_l)$. Synchrotron photons produced in the outer layer with luminosity $L^{\prime\prime}_{\rm syn}$ (as measured in the layer's frame) will be viewed in the spine's frame with energy density $u^\prime_{\rm syn}\approx \Gamma_{\rm rel}^2 \, L^{\prime\prime}_{\rm syn}/4\pi R^2_{l}c \simeq 10^{-4} \, (\Gamma_{\rm rel}/2.5)^2 \, (L^{\prime\prime}_{\rm syn}/10^{43} \ {\rm erg \ s^{-1}})(R_{l}/10^{18} \ {\rm cm})^{-2}$~erg cm$^{-3}$, where $R_{l}$ is the outer layer's radius \citep[for application to neutrino emission from BL Lac objects, see][]{Tavecchio2014, Tavecchio2015}. Assuming that the emission region in our single-zone model is part of the jet spine, then $u^\prime_{\rm syn}$ should be equivalent to $u^\prime_{\rm ext}$. The values of the latter, as obtained from the multi-epoch SED modeling (see Table~\ref{tab:fit}), are comparable to the Doppler-boosted energy density of photons from the outer layer. 
Although the SED of the external radiation field was modeled as a gray body  (see Section~\ref{sec:SED-fits}), power-law energy spectra that are more relevant in the spine-sheath scenario, could equally well describe the SED \citep[for details, see][]{Keivani2018}.
Changes in the outer layer's synchrotron luminosity on a few light crossing times (e.g., due to changes in the dissipation efficiency and/or accretion rate of the black hole), would correspond to month-long variability in the observer's frame, i.e., $\sim R_l/(\delta_l c) \sim 100\, (R_l/10^{18} \ {\rm cm}) (2/\delta_l)$~days. 
Variable external radiation fields on year-long timescales are thus possible in this scenario, which can be consistent with the picture that dissipation physics is more or less similar and the blazar zone is located outside the BLR.  In conclusion, the non-thermal radiation from the sheath of a structured jet is the most likely explanation for the putative external radiation field used in our SED modeling. We caution, however, that our results were obtained under the assumption of isotropic photon fields in the comoving frame of the blob, which breaks down in the structured jet scenario \citep[for relevant discussion, see][]{Reimer2019}. A detailed hybrid leptonic modeling of the SED taking into account anisotropic effects in the $p\gamma$ and photon-photon collisions lies beyond the scope of this paper.

\begin{figure}
    \centering
    \includegraphics[width=0.47\textwidth]{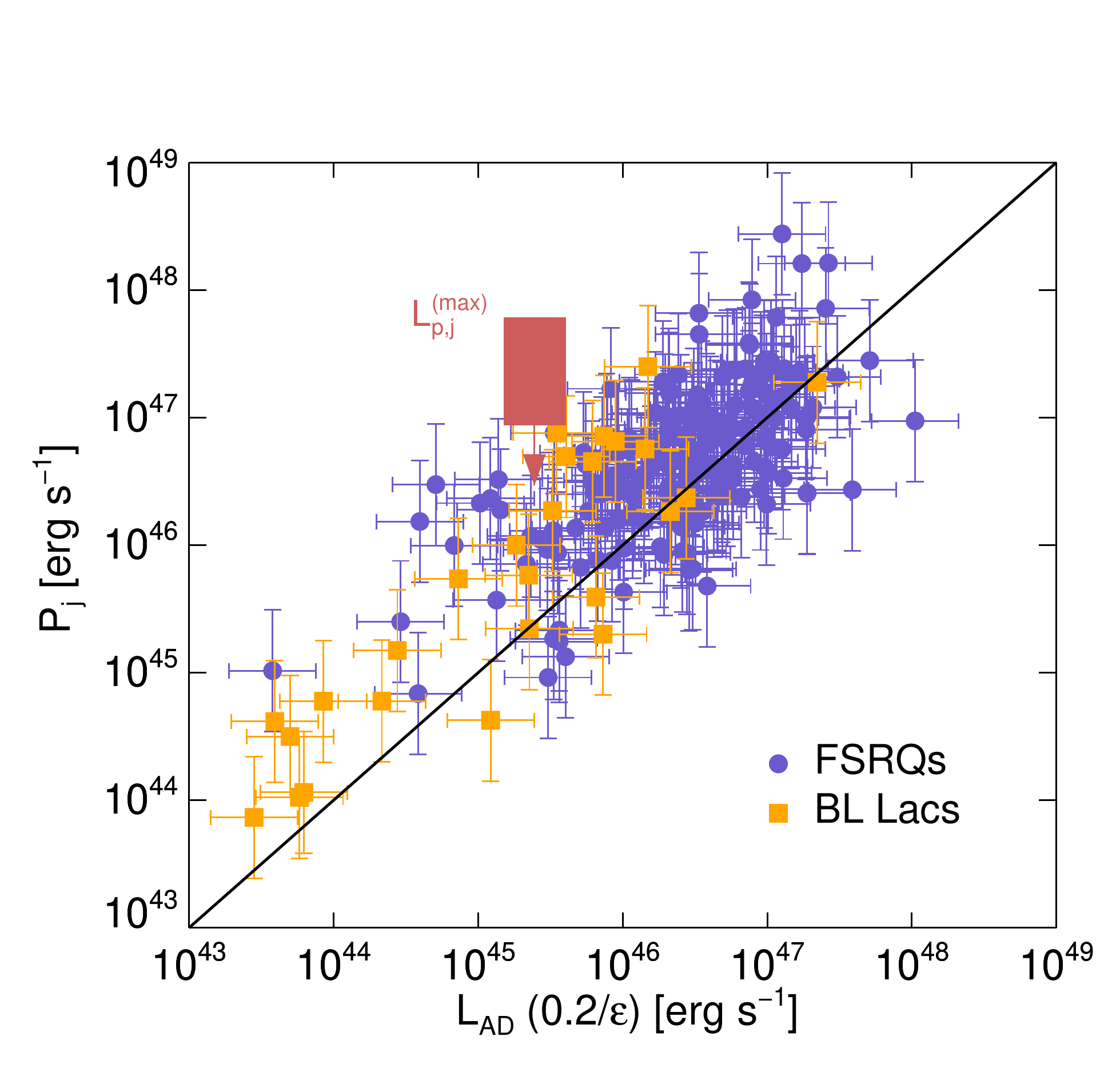}
    \caption{Total jet power estimated through single-zone SED modeling of 217 blazars (colored symbols) versus the accretion power $L_{\rm AD}/\epsilon$, assuming a fixed radiative efficiency $\epsilon=0.2$ \citep[data taken from Table~1 of][]{Ghisellini2014}. The error bars indicate the average uncertainty in the jet power and disk luminosity, as reported by \citet{Ghisellini2014}. The solid line indicates the relation $P_j=\dot{M}c^2$. The colored region denotes the maximal proton jet power inferred from the multi-epoch SED modeling of \txs \, (see Table~\ref{tab:fit}). The width of the box indicates the uncertainty in the disk luminosity \citep{Padovani:2019}.}
    \label{fig:pjet}
\end{figure}

\subsection{Remarks on the absolute jet power}
Finally, we comment on the jet power inferred from our hybrid leptonic interpretation of the SEDs. In general, the jet power can be written as $P_j=\eta_j \dot{M} c^2$, where $\dot{M}$ is the accretion rate onto the black hole and $\eta_j \lesssim 1.5$ is the jet formation efficiency. For  $\eta_j > 1$ the jet power can exceed the accretion power, suggesting efficient extraction of the black hole's rotational energy \citep{Blandford1977}. Such high jet formation efficiencies can be achieved if the black hole is rapidly spinning and there is enough large-scale magnetic flux threading the black hole and accretion flow  \citep{Tchekhovskoy2011, Tchekhovskoy2012, McKinney2012} to lead to the formation of a magnetically arrested disc (MAD) \citep{Bisnovatyi-Kogan1974, Narayan2003}. The accretion disk luminosity can also be written as $L_{\rm AD} = \epsilon \dot{M} c^2$, where $\epsilon$ is the radiative efficiency, which depends on the black hole spin, the disc state (e.g., MAD), and the presence of magnetic winds \citep[e.g.,][]{Tchekhovskoy2011,Avara2016, Morales2018}. For the purposes of this discussion,  we adopt $\epsilon=0.2$\footnote{For very radiatively inefficient flows ($\epsilon \ll 1$), there is no direct proportionality between  $L_{\rm AD}$ and $\dot{M}c^2$. We do not consider this scenario here.}. The (maximum) jet power of \txs \, can be then estimated by $P_{j, \rm  MAD} = \eta_j \, L_{\rm AD}/\epsilon \simeq 6\times 10^{45} \,  (\eta_j/1.5) (0.2/\epsilon) (L_{\rm AD}/8\times10^{44} \, {\rm erg \, s^{-1}})$~erg s$^{-1}$. The maximal jet power inferred from the SED modeling is dominated by the power of relativistic protons, although the latter do not contribute to the blazar's non-thermal emission. We find that $L^{(\max)}_{p,j} \simeq (15-100) \, P_{j, \rm  MAD}$ (see Table~\ref{tab:fit}),  with the highest value  found for epoch 4 being most likely an overestimation due to the poor sampling of the SED. Note also that $L_{p,j}^{(\max)}$ can be reduced by a factor of $\sim 2-3$, if the accelerated protons in the jet have hard power-law energy spectra ($s_p\lesssim 1.5$).

To put things into perspective, Figure~\ref{fig:pjet} shows the total jet power inferred from the leptonic SED modeling of 217 blazars \citep{Ghisellini2014} versus the accretion power for fixed radiative efficiency $\epsilon=0.2$. The range of values for the proton jet power of \txs \, that we obtained in this work (see Table~\ref{tab:fit}) is shown by the colored box. Although the lower upper limit on the proton jet power (which is obtained for the 2017 flare) is consistent with the values inferred from leptonic modeling of gamma-ray blazars, there is  tension with the larger values we obtained. Part of the tension originates from the lack of strong constraints for the SED modeling; note that the highest values of $L^{(\max)}_{p,j}$ are obtained for epochs 3 and 4 where the uncertainty in the X-ray flux is large, and likely are an overestimation of the proton power of the jet. The maximal proton jet power can also be reduced by a factor of $\sim 2-3$, if protons are accelerated into hard power laws (with $s_p\lesssim 1.5$), as illustrated in Figure~\ref{fig:proton-param}. Alternatively, if $\epsilon \ll 0.2$ in \txs, then the inferred accretion power could be much larger, thus releasing part of the tension (the box in Figure~\ref{fig:pjet} would shift horizontally to the right). Finally, an intrinsically lower proton power ($L_{p,j} \ll L^{(\max)}_{p,j}$) would also ease the tension, but would make the prospects for future neutrino detections from \txs \, bleak. 

\section{Summary}\label{sec:summary}
We performed multi-epoch SED modeling of the blazar \txs, including the 2014--2015 period of neutrino flare, within the framework of  a one-zone hybrid leptonic scenario for blazar emission. Having as our baseline the parameter values used to model the 2017 gamma-ray flare in coincidence with \icnu \, \citep[see model LMBB2b in][]{Keivani2018}, we varied as few model parameters as possible (5 out of 11) to derive a theoretical spectrum that describes well the multi-wavelength data of each epoch. Notably, we showed that a time-variable energy density of external photons (within a factor of 5) together with a variable electron injection luminosity (within a factor of 2.5) can explain the observed optical/UV and gamma-ray flux variability in the periods of interest. 

These results suggest that the physical properties of electron acceleration in the jet do not change significantly, and support the external inverse-Compton mechanism as an explanation for the observed gamma-rays~\citep{Keivani2018,Ansoldi2018}. Alternatively, the gamma-rays could be interpreted as synchrotron self-Compton emission from the blazar zone, as proposed for the 2017 flare~\citep[e.g.,][]{Cerruti2019,Gao2019}. While the SSC scenario cannot explain the 2017 flare when the \swift-UVOT and X-shooter data are considered~\citep{Keivani2018}, it remains viable for the 4 epochs considered here. However, it would require changes in almost all model parameters, including $B^\prime$, $R^\prime$, and $\delta$, to account for the large multi-epoch variations of the Compton dominance parameter, i.e., the ratio of the peak inverse Compton and synchrotron fluxes. 

Upon determination of the parameter values needed to explain the multi-epoch SEDs in terms of synchrotron and inverse Compton emissions of primary electrons, we computed the maximal neutrino flux by requiring that any proton-induced emission does not overshoot the X-ray and/or gamma-ray data. We found that the maximal neutrino flux is  better correlated with the X-ray flux of the source, thus confirming the importance of X-ray measurements for constraining the blazar neutrino output in this scenario.

\acknowledgments
M.P. acknowledges support from the Lyman Jr.~Spitzer Postdoctoral Fellowship and NASA Fermi grant No. 80NSSC18K1745.
The work of K.M. is supported by NSF Grant No.~PHY-1620777 and AST-1908689, and the Alfred P. Sloan Foundation. M.S. is supported by NSF Grants No.~PHY-1914579 and PHY-1913607.
A. K. acknowledges support from the Frontiers of Science Fellowship at Columbia University. 
A. T. would like to thank Hans Krimm for useful discussion regarding the BAT Transient Monitor pipeline.
The \textit{Fermi} LAT Collaboration acknowledges generous ongoing support
from a number of agencies and institutes that have supported both the
development and the operation of the LAT as well as scientific data analysis.
These include the National Aeronautics and Space Administration and the
Department of Energy in the United States, the Commissariat \`a l'Energie Atomique
and the Centre National de la Recherche Scientifique / Institut National de Physique
Nucl\'eaire et de Physique des Particules in France, the Agenzia Spaziale Italiana
and the Istituto Nazionale di Fisica Nucleare in Italy, the Ministry of Education,
Culture, Sports, Science and Technology (MEXT), High Energy Accelerator Research
Organization (KEK) and Japan Aerospace Exploration Agency (JAXA) in Japan, and
the K.~A.~Wallenberg Foundation, the Swedish Research Council and the
Swedish National Space Board in Sweden.
 Additional support for science analysis during the operations phase is gratefully
acknowledged from the Istituto Nazionale di Astrofisica in Italy and the Centre
National d'\'Etudes Spatiales in France. This work performed in part under DOE
Contract DE-AC02-76SF00515.

%



\software{Fermipy \citep[]{Wood:2018}, 
Astropy \citep[]{astropy:2013,astropy:2018}, IDL version 8.5  (Exelis Visual Information Solutions, Boulder, Colorado)}.

\appendix
\counterwithin{figure}{section}

\section{The 2014-2015 neutrino flare}\label{sec:appendix-1}
We investigate if the simplest one-zone models of blazar emission can explain the 2014-2015 neutrino flare without violating existing electromagnetic observations, but not necessarily trying to explain the SED \citep[see also][]{Reimer2019}. 

\begin{figure*}
     \centering
     \includegraphics[width=0.55\textwidth, trim=20 20 0 0]{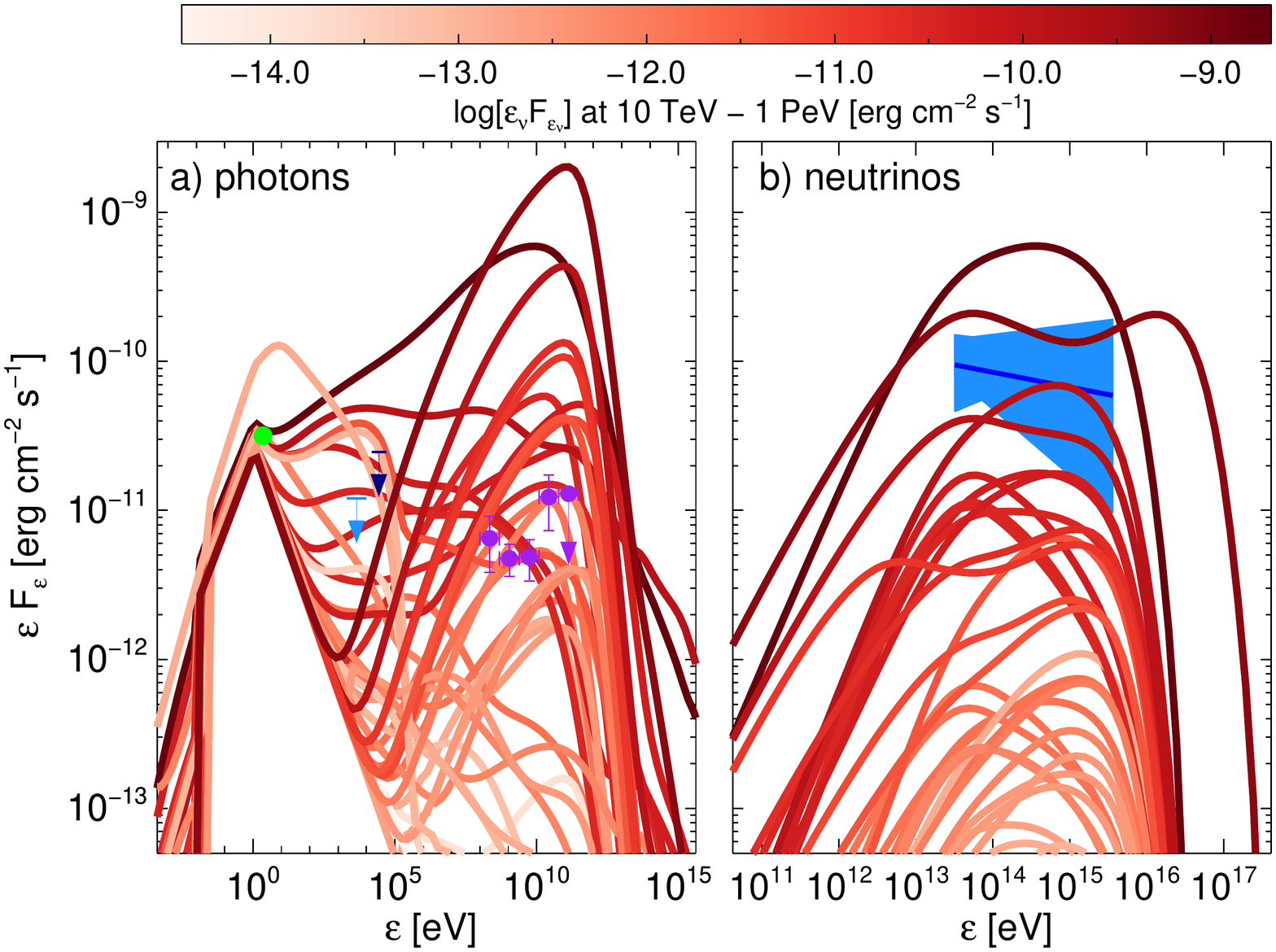}
     \includegraphics[width=0.40\textwidth]{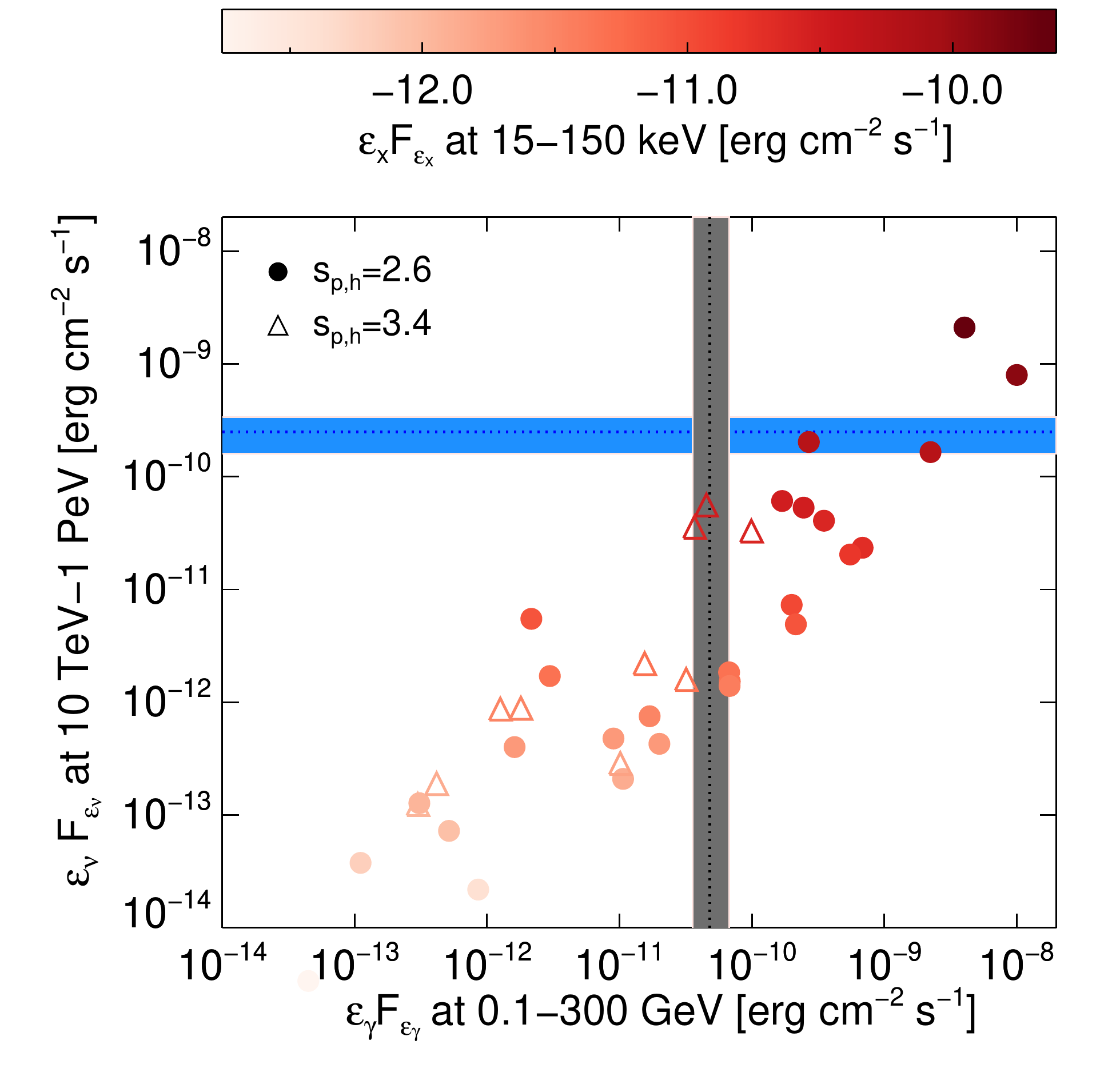}
     \caption{\textit{Left panel:} Broadband photon spectra and all-flavor neutrino spectra for the parameters listed in \tab{param}. The theoretical spectra are colored according to the all-flavor neutrino flux in the 10 TeV -- 1 PeV energy range (see color bar). Overplotted are the \fermi \, spectrum (purple symbols), the X-ray upper limits from \swift-BAT (black arrow) and \maxi \, (blue arrow), and the ASAS-SN optical data point (green filled circle) for epoch 4.  The blue-colored bow tie shows the best-fit all-flavor spectrum (with its 68\% uncertainty region) obtained by IceCube \citep[adopted from Figure 3 of][]{Aartsen2018blazar2}.
     \textit{Right panel:} All-flavor neutrino flux in the 10 TeV-1 PeV energy range plotted against the 0.1-300 GeV energy flux  as predicted by the model for all parameter sets listed in \tab{param}. The color indicates the associated X-ray flux in the 15-150 keV energy range. Results for different proton power-law indices are plotted with different symbols (see inset legend). The horizontal blue-colored stripe indicates the best-fit (with 68\% uncertainties) all-flavor neutrino flux in the 10~TeV$-$1~PeV energy range measured with IceCube \citep{Aartsen2018blazar2}. The best-fit gamma-ray flux  (with 68\% uncertainties) measured by \fermi-LAT  in the 0.1-300 GeV energy range (see Table~\ref{tab:Fermi}) is denoted with the vertical gray-colored stripe. }
     \label{fig:param}
 \end{figure*}

To facilitate the scan of the parameter space, we set the injection of primary electrons to zero and replace the expected synchrotron spectrum by a fixed photon field with a broken power-law energy spectrum that matches the low-energy hump of the archival SED. To reduce the number of free parameters, we do not consider external photon fields. To reduce the energetic requirements as much as possible, we assumed that protons are injected in the source with a broken power-law distribution:
\eqb 
\label{eq:Np}
\dot{N}_{p}(\gamma'_p) & \propto & \left\{ 
\begin{array}{cc}
 \gamma_{p}^{'-s_{p,l}},    &  \gamma'_{p,\min} \le \gamma'_{p}  \le \gamma'_{p,br} \\ \\
 \gamma_p^{'-s_{p,h}} e^{-(\gamma'_p/\gamma'_{p,\max})},    &  \gamma'_{p,br} < \gamma'_{p}  \le \gamma'_{p,\max}
\end{array}
\right.
\eqe 
where we set $s_{p,l}=1.6$, $\gamma'_{p,\min}=1$, $\gamma'_{br}=6.3\times10^4$, and $\gamma'_{p, \max}=6.3\times10^6$, unless stated otherwise. We note that a different choice of values for the aforementioned parameters does not alter our main conclusions.
The total injected luminosity in protons ($L'_p \propto m_p c^2 \int d\gamma'_p \dot{N}_{p}(\gamma'_p) \gamma'_p$) and the high-energy power-law index $s_{p,h}$ are the most important parameters of the proton distribution, as they are  directly related to the flux and slope of the neutrino energy spectrum (for a fixed target photon field). The parameters values used in our search are listed in \tab{param} and our results are presented in Figure \ref{fig:param}. 

We find no single-zone model that can explain the neutrino flare and simultaneously satisfy all the electromagnetic constraints (see left panel in Figure \ref{fig:param}). The broadband photon spectrum, which is a result of the synchrotron and Compton emissions of secondaries produced through photohadronic interactions (photomeson production and Bethe-Heitler pair production processes) and photon-photon pair production, is sensitive to the physical conditions of the source \citep[i.e., magnetic field strength, and photon compactness; for details see][]{Reimer2019}. In the absence of any high-density external photon fields, super-Eddington proton luminosities are required for producing neutrino fluxes comparable to the one measured by IceCube (see \tab{param}). In addition, we find a tight correlation between the 0.1-300 GeV gamma-ray flux and the high-energy neutrino flux (see right panel in Figure \ref{fig:param}), implying a hadronic origin of the gamma-ray emission (to be contrasted with the results of leptonic modeling shown in Figure~\ref{fig:flux-flux}). This tight correlation is also in contrast to the results of \citet{Reimer2019}, where the GeV flux is strongly attenuated by the assumed X-ray external photon field,  which is not included in this treatment.  

\begin{deluxetable*}{ccccc} 
\tablecaption{Values used in the parameter space search for the neutrino emission of the 2014--2015 flare. \label{tab:param}}
\tablecolumns{5}
\tablenum{1}
\tablewidth{0pt}
\tablehead{
\colhead{Magnetic field} & \colhead{Blob radius} &  \colhead{Doppler factor} & \colhead{Power-law index} & \colhead{Proton jet power\tablenotemark{a}} \\
$B'$ [G] & $R'$ [cm] & $\delta$ & $s_{p,h}$ & $L_{p,j}$ [erg s$^{-1}$]  
}
\startdata 
0.01 & $10^{16}$ & 10 & 2.6 &  $2.5\times10^{49}$ \\
0.01 & $10^{16}$ & 10 & 2.6 &  $8\times10^{49}$ \\
0.01 & $10^{16}$ & 10 & 2.6 &  $2.5\times10^{50}$ \\
0.01 & $10^{16}$ & 10 & 2.6 &  $8\times10^{50}$ \\
0.01 & $10^{16}$ & 10 & 2.6 &  $8\times10^{50}$\tablenotemark{\dag} \\
0.01 & $10^{16}$ & 10 & 2.6 &  $8\times10^{50}$\tablenotemark{\ddag} \\
0.1  & $10^{17}$ & 10 & 2.6 & $4\times10^{50}$ \\
0.1  & $10^{17}$ & 10 & 2.6 & $1.3\times10^{51}$ \\
0.1  & $10^{17}$ & 10 & 3.4 & $4\times10^{50}$ \\
0.1  & $10^{17}$ & 10 & 3.4 & $1.3\times10^{51}$ \\
0.1  & $10^{17}$ & 20 & 2.6 & $1.6\times 10^{51}$\\
0.1  & $10^{17}$ & 20 & 3.4 & $1.6\times 10^{51}$\\
0.1  & $10^{16}$ & 20 & 2.6 & $10^{50}$\\
0.1  & $10^{16}$ & 20 & 2.6 & $10^{51}$\\
0.1  & $10^{15}$ & 20 & 2.6 & $10^{49}$\\
0.1  & $10^{15}$ & 20 & 2.6 & $10^{50}$\\
1    & $10^{17}$ & 10 &  2.6 & $4\times10^{49}$ \\
1    & $10^{17}$ & 10 &  3.4 &  $4\times10^{50}$\\
1    & $10^{16}$  & 10 & 2.6 &  $4\times10^{47}$\\
1    & $10^{16}$  & 10 &  2.6 & $4\times10^{48}$\\
1    & $10^{15}$  & 10 &  2.6 & $4\times10^{47}$\\
1    & $10^{15}$  & 10 &  2.6 & $1.3\times10^{49}$\\
10   & $10^{17}$ & 10 & 3.4 & $4\times10^{49}$ \\
10  & $10^{16}$ & 10 & 3.4 & $4\times10^{48}$\\
10  & $10^{16}$ & 10 & 3.4 & $4\times10^{49}$ \\
30  & $10^{17}$ & 10 & 2.6  & $4\times10^{47}$\\
30  & $10^{16}$ & 10 & 2.6  & $4\times10^{46}$\\
30  & $10^{16}$ & 10 & 2.6  & $4\times10^{47}$\\
30  & $10^{16}$ & 10 & 2.6  & $4\times10^{48}$\\
30 & $10^{17}$ & 20 & 2.6  & $1.6\times10^{48}$\\
30 & $10^{17}$ & 20 & 3.4  & $1.6\times10^{49}$\\
30 & $10^{15}$ & 20 & 2.6 & $1.6\times10^{48}$\\
\enddata
\tablenotetext{a}{This is a derived quantity, defined as $L_{p,j}=2\pi R^{'2}c\Gamma^2u'_p$, where $\Gamma \approx \delta/2$.}
\tablenotetext{\dag}{$\gamma'_{p,br}=6.3\times10^3$.}
\tablenotetext{\ddag}{$\gamma'_{p,\max}=6.3\times10^7$.}
\end{deluxetable*}

\bibliographystyle{aasjournal} 
\bibliography{txs.bib}
\end{document}